\documentclass[twocolumn]{openjournal}
\usepackage{amsmath}
\usepackage{tabularx}
\usepackage{booktabs}

\usepackage{xcolor}
\usepackage{textgreek}
\usepackage[T1]{fontenc}
\usepackage[utf8]{inputenc}
\usepackage[english]{babel}

\usepackage{hyperref}
\hypersetup{
    unicode, 
    colorlinks=true,
    linkcolor=linkcolor,
    citecolor=linkcolor,
    filecolor=linkcolor,
    urlcolor=linkcolor,
}
\usepackage{color,colortbl}
\definecolor{linkcolor}{rgb}{0.0,0.0,1.0}
\usepackage{tensind}
\tensordelimiter{?}
\DeclareGraphicsExtensions{.bmp,.png,.jpg,.pdf}
\usepackage{verbatim}
\usepackage[normalem]{ulem}
\usepackage{orcidlink}
\usepackage{soul}

\newcommand{\MBH}{M_{\rm BH}}

\newcommand{\lrb}[1]{\left({#1}\right)}
\newcommand{\lrsb}[1]{\left[{#1}\right]}
\newcommand{\lra}[1]{\left\langle{#1}\right\rangle}
\newcommand{\abs}[1]{\left|#1\right|}

\begin{document}
\title{Long-Term Evolution of Hills-Captured Stars with TDE Disk Collisions: \\ 
 Links to Repeating Nuclear Transients}

\author{Howard Hao-Tse Huang$^{1,\star,\dagger}$\orcidlink{0000-0003-4647-2591}}
\altaffiliation{$^\star$These authors contributed equally to this work.}
\email[$^{\dagger}$]{haotse813@berkeley.edu}

\author{Philippe Z. Yao$^{2,3,\star,\ddagger}$\orcidlink{0000-0003-3024-7218}}
\email[$^\ddagger$]{philippe.yao@columbia.edu}

\author{Wenbin Lu$^1$\orcidlink{0000-0002-1568-7461}}

\author{Eliot Quataert$^3$\orcidlink{0000-0001-9185-5044}}

\affiliation{$^1$Department of Astronomy, University of California at Berkeley, Berkeley, CA 94720, USA}
\affiliation{$^2$Department of Physics and Columbia Astrophysics Laboratory, Columbia University, New York, NY 10027, USA}
\affiliation{$^3$Department of Astrophysical Sciences, Princeton University, Princeton, NJ 08544, USA}

\begin{abstract}
Quasi-periodic eruptions (QPEs) and repeating partial tidal disruption event (rpTDE) candidates 
are widely interpreted as being powered by a star on an hours-to-years orbit 
around a massive black hole (MBH).  Every such orbiter model presupposes a star delivered onto an extraordinarily compact orbit without being destroyed, but it is unclear whether this is achievable. We follow the coupled orbital and stellar evolution of stars captured by the Hills mechanism, combining gas drag and per-collision mass loss from recurrent TDE disks with gravitational-wave (GW) orbital decay and stochastic two-body relaxation in a Monte Carlo model. Conventionally, the EMRI-formation criterion is based on comparing the timescales of GW decay and orbital angular momentum relaxation. We find that this criterion severely overestimates the EMRI rate: for  relaxation timescales shorter than a Hubble time, most Hills-captured stars are tidally disrupted before GW emission can circularize the stellar orbit. Instead, TDE-disk drag plays a crucial role in the orbital evolution and can boost the fraction of stars reaching QPE orbits by an order of magnitude or more, assuming that the disk aspect ratio is relatively large $H/r \sim 0.1$. We also find that collisional ablation imposes a minimum orbital period that rises with MBH mass and precludes sub-day orbits for $M_{\rm BH} \gtrsim 10^6\,M_\odot$, unless GWs alone can drive orbital decay faster than orbital relaxation. This can explain the observed preference of short-period QPEs for low-mass MBHs. Most known rpTDEs lie below the Hills-capture period minimum and thus require additional orbital decay, which TDE-disk drag can plausibly supply. These results support the viability of the orbiter model; our predictions of the MBH mass and orbital period distributions can be used to further test this model.
\end{abstract}

% Write your keywords here
% \begin{keywords}
%     {First, second}
% \end{keywords}

\maketitle

\section{Introduction}
Quasi-periodic eruptions (QPEs) are luminous soft X-ray flares that repeat every few hours to days in the nuclei of low-mass galaxies \citep{Miniutti2019, Giustini2020, Arcodia2021, Chakraborty2021, Arcodia2022, Miniutti2023, Arcodia2024a, Arcodia2025, Hernandez-Garcia2025a}. Individual eruptions last $\sim 10$--$20\%$ of the recurrence time, reach blackbody temperatures of $kT \sim 50$--$200 \,\rm eV$ and bolometric luminosities of $L_{\rm bol} \sim 10^{41}$--$10^{44}\,\rm erg\,s^{-1}$, and stand out by an order of magnitude or more against a cooler quiescent level. Black hole mass estimates for the hosts consistently favor the low-mass end of the massive black hole (MBH) population, $M_{\rm BH} \sim 10^5$--$10^7\,M_\odot$ \citep{Wevers2022}. Roughly half of the known sources were identified in blind X-ray searches \citep{Miniutti2019, Giustini2020, Chakraborty2021, Arcodia2021, Arcodia2024a, Arcodia2025}; the remainder surfaced through archival mining or targeted X-ray monitoring of longer-duration nuclear transients, in several cases optically selected and spectroscopically confirmed tidal disruption events (TDEs) \citep{Quintin2023, Bykov2025, Nicholl2024, Hernandez-Garcia2025a, Chakraborty2025, Baldini2026}.

Several lines of timing evidence point to an underlying orbital clock. The alternation between long and short recurrence intervals seen in a few sources \citep{Miniutti2019, Giustini2020, Arcodia2022, Arcodia2024a} arises naturally if a body on a mildly eccentric orbit intersects the plane of an accretion disk twice each orbit. Although disk-instability scenarios remain under consideration for parts of the phenomenology \citep{Miniutti2019, Sniegowska2020, Raj2021, Pan2022, Kaur2023}, most recent theoretical work adopts this orbiter picture, in which a star or compact object periodically interacts with gas in the innermost nucleus \citep[e.g.,][]{Xian2021, Sukova2021, Krolik2022, Zhao2022, Metzger2022, Linial2023, Lu2023, Tagawa2023, Franchini2023, Linial2024b, Yao2025}. The gas itself is most plausibly the relic of a TDE: disruptions recur in a given galaxy at a rate $\Gamma_{\rm TDE} \sim 10^{-5}$--$10^{-4}\,\rm yr^{-1}\,gal^{-1}$ \citep[e.g.,][]{Stone2016, Gezari2021, Yao2023, Shepherd2026}, and \citet{Linial2023} showed that the resulting disks, once viscously spread, generically engulf the orbits of the most tightly bound stars. Consistent with this expectation, some QPEs have now been observed to switch on a few years after spectroscopically confirmed TDEs, most notably AT 2019qiz \citep{Nicholl2024}, with a growing list of similar associations \citep{Quintin2023, Bykov2025, Chakraborty2025}. %The prototypical source GSN 069 likewise shows long-term X-ray evolution consistent with a long-lived TDE disk \citep{Miniutti2019, Giustini2020, Miniutti2023}, and star--disk impact models reproduce the observed QPE energetics and temperatures when the impacted disk is TDE-supplied \citep{Franchini2023}.

Every version of the orbiter model \citep[e.g.,][]{Xian2021, Krolik2022, Lu2023, Linial2023, Franchini2023, Linial2025} rests on the same demanding prerequisite: a star that has survived on an extraordinarily compact orbit. Matching the observed recurrence times requires orbital periods of hours to days, corresponding to semimajor axes of tens to hundreds of gravitational radii, and timing solutions favor low to moderate eccentricities \citep[e.g.,][]{Franchini2023, Zhou2024}. Stars neither form on such orbits nor scatter onto them intact: the required orbital binding energy exceeds that of typical nuclear-cluster stars by orders of magnitude, so a dissipative process must deliver the star and circularize it without destroying it along the way. The stakes are raised by the short QPE-emitting lifetimes implied by collisional mass loss, $\lesssim 10^3$ yr \citep{Linial2023, Yao2025}: sustaining the observed QPE abundance \citep{Arcodia2024b} then requires the delivery channel to operate frequently in ordinary low-mass nuclei. The long-term orbital evolution of stars near MBHs determines whether QPE systems can exist at all, which black hole masses can host them, what orbital periods are attainable, and how QPE activity should correlate with the TDE history of the nucleus. Quantifying this evolution, and thereby testing the viability of the orbiter model end to end, is the goal of this paper.

The natural starting point is the Hills mechanism \citep{Hills1988}. A stellar binary deflected onto a nearly radial orbit is tidally separated upon crossing its binary tidal radius; one component is ejected as a hypervelocity star \citep{Yu2003, Brown2015}, while its companion is captured onto a bound orbit with semimajor axis $a_0 \simeq (a_{\rm b}/2)(M_{\rm BH}/M_{\rm b})^{2/3}$ and eccentricity $1 - e_0 \simeq (M_{\rm b}/M_{\rm BH})^{1/3} \approx 0.005$--$0.03$, where $a_{\rm b}$ and $M_{\rm b}$ are the binary separation and mass \citep{Hills1988, Sari2010}. For solar-type binaries around a $10^6\,M_\odot$ MBH, this yields $e \sim 0.97-0.995$, orbital periods of months to decades, and pericenters at a few stellar tidal radii. The post-capture orbit thus has the right pericenter for an eventual QPE system; reducing the orbital periods into hours to days requires increasing the orbital binding energy by two to three orders of magnitude while keeping the pericenter outside the tidal disruption radius.

Gravitational wave (GW) emission \citep{Peters1964} is the textbook route. Because GW losses are concentrated near pericenter, they shrink and circularize the orbit at nearly constant pericenter distance, converting a Hills-captured star into a stellar extreme mass-ratio inspiral (EMRI) --- the same pathway long studied for compact-object EMRIs \citep[e.g.,][]{HopmanAlexander2005, AmaroSeoane2018}. The competition is two-body relaxation: for a nearly radial orbit, the angular momentum random-walks on a timescale $t_{L,\rm rel} \sim (r_{\rm p}/a)\,t_{\rm 2B,rel}$, far shorter than the two-body relaxation time $t_{\rm 2B,rel}$ \citep{FrankRees1976, LightmanShapiro1977}, so a star whose pericenter already lies at a few tidal radii is perpetually at risk of diffusing into the loss cone and being disrupted, the very process that supplies TDEs in the first place \citep[e.g.,][]{Stone2016}. EMRI formation is conventionally deemed successful when the GW decay timescale beats the angular momentum relaxation timescale \citep{HopmanAlexander2005}, and QPE formation scenarios have largely inherited this criterion \citep[e.g.,][]{Linial2023, Lu2023}. However, we show in this paper that this criterion can substantially overestimate the surviving fraction when the two-body relaxation time is short.

Gas provides a faster --- and, in QPE hosts, recurrently available --- source of dissipation. Star--disk interactions have long been studied as a mechanism for capturing and grinding down stellar orbits in active nuclei \citep{Syer1991, Rauch1995, MacLeod2020}, and accretion disks can boost EMRI formation rates by orders of magnitude over the relaxation-only channel \citep{PanYang2021}. QPE hosts lack persistent AGN disks, but each TDE leaves behind up to a few tenths of a solar mass of bound gas that circularizes near the tidal radius and spreads viscously over the following years to decades \citep{LyndenBell1974, Cannizzo1990, ShenMatzner2014, vanVelzen2019, Mummery2020, 2022hxga.book....3L, 2025ApJ...985...77P, Winter2026}, with late-time spectral modeling indicating disks that grow to roughly ten times the circularization radius \citep{Guolo2025}. These disks need not be geometrically thin: magnetic pressure from MRI-amplified toroidal fields and advected vertical flux can elevate them to $H/R \sim 0.1$--$1$, keeping them thicker and less dense than standard $\alpha$-disks even as $\dot{M}$ declines \citep[e.g.,][]{Begelman2007, Jiang2014, Sadowski2016, Begelman2017, Curd2019}. A Hills-captured star with a pericenter of a few tidal radii is therefore episodically immersed in fresh disks at the TDE rate. Each active phase produces collisions whose drag removes orbital energy and shrinks the orbit \citep{Linial2024}; the associated period decay may already be observed in some repeating nuclear transients \citep{Payne2021, Linial2024}. The same collisions, however, ablate the star: ram pressure strips the envelope, and repeated shock heating inflates it, enhancing subsequent mass loss \citep{Yao2025}. Tidal heating may impose further limits during the final stages of inspiral \citep{Li2013, 2021MNRAS.503..603L, Yao2026}.

% In this paper we follow the coupled orbital and stellar evolution of Hills-captured stars from capture until destruction or arrival on a QPE-like orbit. Our framework combines (i) gas drag from collisions with recurrent, viscously spreading TDE disks; (ii) per-collision mass loss calibrated to the hydrodynamic simulations of \citet{Yao2025}, coupled to the star's mass--radius response computed with \texttt{MESA} \citep{Paxton2011}; (iii) GW orbital decay \citep{Peters1964}; and (iv) stochastic two-body relaxation of the orbital angular momentum, implemented within a Monte Carlo model. We use this framework to address three questions that together constitute a quantitative viability test of the orbiter model: (1) what is the minimum orbital period a star can reach before collisional mass loss destroys it; (2) over what range of black hole masses and orbits does disk drag, rather than GW emission, control the circularization; and (3) what fraction of Hills-captured stars evades tidal disruption and reaches hour-to-day period orbits, as a function of the TDE rate, the relaxation time, and $M_{\rm BH}$? 

In this paper, we follow the coupled orbital and stellar evolution of Hills-captured stars under disk collisions, GW emission, and two-body relaxation. In \S\ref{sec:gas-drag-mass-loss} we construct the model ingredients: the viscous spreading of recurrent TDE disks, the orbit-averaged collisional drag, the simulation-calibrated mass loss, and the associated stellar radius response. \S\ref{sec:disk-only} isolates disk collisions and derives the period floor imposed by collisional mass loss, and \S\ref{sec:disk-GW-model} adds GW orbital decay and delineates where each mechanism dominates. In \S\ref{sec:full-model} we introduce two-body relaxation via Monte Carlo simulations, revisit the conventional EMRI-formation criterion, and compute the fraction of Hills-captured stars that reach QPE orbits across the model parameter space. We summarize our findings and discuss their observational implications in \S\ref{sec:discussion}.

\section{Star--Disk Collisions: Mass Loss and Disk Drag}\label{sec:gas-drag-mass-loss}

Star--disk collisions produce two coupled effects: \emph{mass loss}, as the stellar envelope is ablated by the disk's ram pressure, and \emph{disk drag}, as momentum exchange with the disk removes orbital energy and angular momentum. Mass loss and disk drag operate simultaneously and in direct competition: the former limits the stellar lifetime before depletion, whereas the latter helps dissipate the energy and shrink the orbit. Understanding the dynamics between these two opposing effects is pivotal in assessing the feasibility of QPE orbit formation in Hills-breakup stars.

In this section, we lay the foundations for our subsequent simulations:  a simplified time-averaged TDE disk model, a mass loss and gas drag prescription based on \citet{Yao2025} and \citet{Linial2024}, and a \texttt{MESA} model for the stellar response to mass loss.

\subsection{Spreading Disk Model}\label{sec:disk-model}
We consider a viscously spreading disk following a TDE. The disk has mass $M_d$ and radius $r_d$, with initial values $M_{d,0}$ and $r_{d,0}$. The spreading of the disk is determined by a one-zone model with $\alpha$ viscosity prescription and constant disk height ratio $h$. At any given moment of time, the radial structure of the disk is fixed by the uniform accretion rate. The effect of inner boundary is ignored in our model.

The disk spreads according to the viscous timescale $t_{\text{vis}}$:
\begin{equation}
    \frac{1}{r_d}\frac{dr_d}{dt} = \frac{1}{t_{\text{vis}}}.
\end{equation}
For a disk with constant disk aspect ratio $h$ and viscosity parameter $\alpha$, $t_{\text{vis}}$ is given by
\begin{equation}
    t_{\text{vis}} = \frac{1}{\alpha h^2 \Omega_K},
\end{equation}
where $\Omega_K=\sqrt{G\MBH /r_d^3}$ is the Keplerian angular frequency at the disk radius. In this paper the value of $\alpha$ is fixed to 0.1. The disk radius expands due to the viscosity:
\begin{equation}
    r_d = r_{d,0}\lrb{\frac{t}{t_{\rm vis,0}}}^{2/3},
\end{equation}
where $t_{\rm vis,0}=t_{\text{vis}}(r_d=r_{d,0})$ is the initial viscous time.
With no outflow and conservation of the angular momentum in the disk ($J_d=M_d\sqrt{G\MBH r_d}$), the disk mass can be expressed as
\begin{equation}
    M_d = \frac{J_d}{\sqrt{G\MBH r_d}} = M_{d,0}\lrb{\frac{t}{t_{\rm vis,0}}}^{-1/3}.
\end{equation}

The surface density profile $\Sigma(r)$ of the disk at any given time is determined by the requirement of the uniform accretion rate $\dot{M}_d$ at all radii:
\begin{equation}
    \frac{\Sigma r^2}{t_{\text{vis}}(r)} \sim \dot{M}_d.
\end{equation}
This gives the spatial dependence of surface density $\Sigma \propto r^{-1/2}$. Combined with the constraint of the total disk mass, this sets the surface density profile:
\begin{equation}
    \Sigma(r,t) =
     \begin{cases}
         \Sigma_{d,0}\lrb{\frac{t}{t_{\rm vis,0}}}^{-4/3}\lrb{\frac{r}{r_{d,0}}}^{-1/2},&\text{for }r<r_d,\\
         0,&\text{for }r>r_d,
     \end{cases}
\end{equation}
where $\Sigma_{d,0}=3M_{d,0}/(4\pi r_{d,0}^{2})$. 

We next consider the initial conditions of the disk. We assume the disruption of a $1\,M_\odot$ MS star and ignore the potentially deep-penetrating TDE. The initial disk radius is equal to the circularization radius
\begin{equation}
    r_{d,0} = 2\,r_{t,\odot},
\end{equation}
where $r_{t,\odot} = R_{\odot}\lrb{\MBH / M_{\odot}}^{1/3}$ is the tidal radius.
The initial disk mass $M_{d,0}$ is determined by the amount of material that circularizes and assembles the disk.
Without outflow, half of the stellar mass will remain bound to the MBH and form the disk, with $M_{d,0}=0.5\,M_\odot$. However, the above picture is complicated by the super-Eddington luminosity at early times.
In TDEs, the fallback rate after its peak can be modeled as
\begin{equation}
    \dot{M}_{\rm fb} = \frac{M_\odot}{4\,t_{\rm peak}}\lrb{\frac{t}{t_{\rm peak}}}^{-5/3},
\end{equation}
where $t_{\rm peak}$ is the time of the peak fallback rate, given by \citep{Bandopadhyay2024}:
\begin{equation}
    t_{\rm peak} = 30\,{\rm days} \lrb{\frac{\MBH}{10^6\,M_\odot}}^{1/2}.
\end{equation}
For rapid circularization, the luminosity associated with the fallback is
\begin{equation}
    L_{\rm fb} = \frac{G\MBH \dot{M}_{\rm fb}}{2\,r_{d,0}}.
\end{equation}
At early times, $L_{\rm fb}$ can be larger than the Eddington luminosity
\begin{equation}
    L_{\rm edd} = 1.3\times10^{44}\,{\rm erg}\,{\rm s}^{-1}\lrb{\frac{\MBH}{10^6\,M_\odot}}.
\end{equation}
When $L_{\rm fb}>L_{\rm edd}$, we expect significant outflow, and the accretion rate onto the disk is throttled to
\begin{equation}
    \dot{M}_{\rm edd} = \frac{L_{\rm edd}}{G\MBH/(2r_{d,0})}.
\end{equation}
This leads to a decrease in $M_{d,0}$:
\begin{align}\label{eq:M-d0}
    M_{d,0} &= \min\left(0.5\,M_\odot, \dot{M}_{\rm edd}t_{\rm edd}+\int_{t_{\rm edd}}^\infty \dot{M}_{\rm fb}dt\right)\nonumber\\
    &=
    \begin{cases}
        0.28\,M_\odot \lrb{\frac{\MBH}{10^6 M_\odot}}^{1/3}, & \MBH < 5.6\times10^6 M_\odot, \\
        0.5\,M_\odot, & \MBH > 5.6\times10^6 M_\odot.
    \end{cases}
\end{align}
In the above, $t_{\rm edd}$ is the time when $L_{\rm fb}=L_{\rm edd}$, and $\min(0.5\,M_\odot,\cdots)$ is to account for our inaccurate modeling of $\dot{M}_{\rm fb}$ before $t_{\rm peak}$.
The super-Eddington fallback affects systems with $M_{\rm BH}\lesssim 5.6\times 10^6\,M_{\odot}$, and $M_{d,0}$ can drop to as low as $0.13\,M_{\odot}$ for $M_{\rm BH}=10^5 \,M_\odot$.

Since the stellar orbital period ($\lesssim\,1\,{\rm yr}$) is much shorter than the TDE disk lifetime ($\gtrsim10\,{\rm yr}$), we can construct the time-averaged surface density profile $\lra{\Sigma}$ over multiple TDEs. Assuming the TDE disk is assembled at $t_{\rm vis,0}$ and viscously spreads until time $t_f\gg t_{\rm vis,0}$, $\lra{\Sigma}$ has the following expression:
\begin{align}
    \lra{\Sigma}(r)&=\Gamma_{\rm TDE}\int_{t_{\rm vis,0}}^{t_f}\Sigma(r,t)\,dt \nonumber\\
    &\simeq\lra{\Sigma}_0\times
    \begin{cases}
        \left(\dfrac{r}{r_{d,0}}\right)^{-1/2}, & r<r_{d,0},\\
        \left(\dfrac{r}{r_{d,0}}\right)^{-1}, & r_{d,0}<r<r_d(t_f),\\
        0, & r>r_d(t_f),
    \end{cases}
\end{align}
where $\Gamma_{\rm TDE}$ is the TDE rate per galaxy and $\lra{\Sigma}_0$ is the normalization:
\begin{align}
    \lra{\Sigma}_0
    &=3\Sigma_{d,0}\Gamma_{\rm TDE}t_{\rm vis,0}\nonumber\\
    &=53\,{\rm g}\,{\rm cm}^{-2}
    \lrb{\frac{M_{d,0}}{0.5\,M_\odot}}
    \lrb{\frac{h}{0.1}}^{-2}\nonumber\\
    &\quad\times
    \lrb{\frac{\Gamma_{\rm TDE}}{10^{-4}\,{\rm yr}^{-1}}}
    \lrb{\frac{\MBH}{10^6\,M_\odot}}^{-2/3}.
\end{align}
To accurately set the maximum extent of the disk $r_d(t_f)$ would require the unknown disk lifetime $t_f$. Here we conservatively set $r_d(t_f)=10\,r_{d,0}$ based on the late time TDE disk SED modeling \citep{Guolo2025}. We emphasize that the uncertain $t_f$ only affects the averaged disk profile in its outer edge $r_d(t_f)$, since $\Sigma \propto t^{-4/3}$ at a given radius and most of the contribution to $\lra{\Sigma}$ comes from the early time $t\sim t_{\rm vis,0}$.

In summary, for a given $\MBH$, our simplified time-averaged disk model has a fixed initial disk mass $M_{d,0}$ (Eq.~\ref{eq:M-d0}) and two free parameters: (1) the disk height ratio $h$, and (2) the TDE rate in the galaxy $\Gamma_{\rm TDE}$.

\subsection{Gas drag}
The gas drag and mass loss from a single star--disk collision depend on the collision speed $v_{\rm col}$, the stellar mass $M_*$ and radius $R_*$, and the local disk properties. In general, the gas drag depends on the relative velocity between the star and the disk gas, and therefore on the orientation of the stellar orbit relative to the disk plane. To approximately average over different orbital orientations, we treat the gas drag as an instantaneous velocity change directed opposite to the star's velocity. The star follows a Keplerian orbit with semi-major axis (SMA) $a$ and eccentricity $e$. For simplicity, we assume that the collisions occur at pericenter and, if the disk extends sufficiently far, at apocenter.

For a star that collides with the disk at a speed much higher than the escape speed on the stellar surface, the drag force is geometric and has the following expression \citep{Linial2024}:
\begin{equation}
    F_{\rm drag}=\frac{1}{2}\rho v_{\rm col}^2\cdot\pi R_*^2,
\end{equation}
where $\rho$ is the local disk density. The collision speed is set to be the stellar velocity and is given by
\begin{align}
    v_{\rm col}&=\sqrt{G\MBH\lrb{\frac{2}{r}-\frac{1}{a}}}\nonumber\\
    &=\begin{cases}
        \sqrt{G\MBH(1+e)/r_p},&\text{for }r=r_p, \\
        \sqrt{G\MBH(1-e)/r_a},&\text{for }r=r_a,
    \end{cases}
\end{align}
where $r_p=a(1-e)$ and $r_a=a(1+e)$ are the pericenter and apocenter radii.
Integrating the above drag force over the entire disk column gives the star's energy loss from drag per collision:
\begin{equation}\label{eq:Delta-E-drag}
    \lrb{\Delta E}_{\rm drag}=-\int F_{\rm drag}dz=-\frac{1}{2}\lra{\Sigma}(r)v_{\rm col}^2\cdot\pi R_*^2.
\end{equation}
Our use of the time-averaged disk profile is justified because $\lrb{\Delta E}_{\rm drag}\propto\Sigma$. For a single collision, $|\lrb{\Delta E}_{\rm drag}|$ is small compared to the magnitude of orbital energy $E=-G\MBH M_*/(2a)$, and the corresponding change in $a$ is
\begin{equation}\label{eq:a-drag}
    \frac{\lrb{\Delta a}_{\rm drag}}{a}=-\frac{\lrb{\Delta E}_{\rm drag}}{E}.
\end{equation}
Under our simplifying assumptions that the gas drag acts instantaneously opposite to the stellar velocity and that collisions occur only at pericenter or apocenter, a collision at pericenter preserves the pericenter radius. In contrast, a collision at apocenter also changes $r_p$:
\begin{equation}\label{eq:rp-drag}
    \lrb{\Delta r_p}_{\rm drag}=\begin{cases}
        0,&\text{for }r=r_p \\
        2\lrb{\Delta a}_{\rm drag},&\text{for }r=r_a
    \end{cases}.
\end{equation}

\subsection{Mass loss and stellar response}\label{sec:mass-loss}
A star--disk collision also induces stellar mass loss and structural change. The 2D Athena++ hydrodynamic simulations of \citet{Yao2025} show that the mass loss per collision is set primarily by the disk ram pressure and scale height, and can be expressed in the following analytic approximation:
\begin{equation}\label{eq:mass-loss-general}
    \frac{\Delta M_*}{M_*}=-f_{\rm ml}\lrb{\frac{H}{R_*}}\lrb{\frac{p_{\rm ram}}{p_*}},
\end{equation}
where $f_{\rm ml}$ is a dimensionless factor that determines the mass loss efficiency, $H=h\cdot r$ is the local disk height at the point of collision, $p_{\rm ram}$ is the peak ram pressure experienced by the star during collision, and $p_*=GM_*^2/(4\pi R_*^4)$ is the average pressure within the star.
For consistency, we follow the conventions in \citet{Yao2025} and define $H$ to be the half-width-half-maximum of the disk's Gaussian density profile in the vertical direction.
In this definition, $p_{\rm ram}$ has the following expression:
\begin{equation}
    p_{\rm ram}=\frac{1}{2}\rho_{\rm d}v_{\rm col}^2=\frac{1}{2}\lrb{\sqrt{\frac{\ln2}{\pi}}\frac{\lra{\Sigma}}{H}}v_{\rm col}^2,
\end{equation}
where $\rho_{\rm d}$ is the midplane density.
Similarly to the case of disk drag, our use of averaged surface density $\lra{\Sigma}$ here is justified because $\Delta M_*\propto p_{\rm ram}\propto\Sigma$.

\begin{figure}
\centering
\includegraphics[width=\columnwidth]{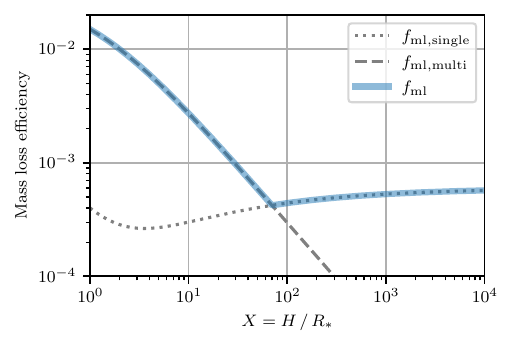}
\caption{The dimensionless mass loss efficiency factor $f_{\rm ml}$. The gray dotted and dashed lines show the efficiency in single and multi shock regimes (Eq.~\ref{eq:f-ml-two-regimes}), and the blue line shows the maximum that we use in the paper (Eq.~\ref{eq:f-ml}).}
\label{fig:f-ml}
\end{figure}

In \citet{Yao2025}, two regimes of collisions are identified depending on whether the thermal time of the shocked envelope, $t_{\rm th}$, is short or long compared to the stellar orbital period $P$ (``single'' vs.\ ``multiple'' shock regimes). Each regime has its own mass loss efficiency \citep[Eq.~4 of][]{Yao2025}:
\begin{align}\label{eq:f-ml-two-regimes}
    f_{\rm ml,single}&=10^{-3}\cdot\frac{X^{0.25}-1.2X^{0.6}+0.6X}{X},\nonumber\\
    f_{\rm ml,multi}&=\frac{0.03}{1+X},
\end{align}
where $X=H/R_*$ is the disk scale height in units of stellar radius.
Without modeling the detailed long-term evolution of the stellar structure, we do not attempt to distinguish these two regimes; instead, we adopt a conservative treatment and set the mass loss per collision to be the maximum of the two regimes:
\begin{equation}\label{eq:f-ml}
    f_{\rm ml}=\max\lrb{f_{\rm ml,single},f_{\rm ml,multi}}.
\end{equation}
Figure~\ref{fig:f-ml} shows $f_{\rm ml,single}, f_{\rm ml,multi},$ and $f_{\rm ml}$ as functions of $X=H/R_*$. The multi-shock case dominates at smaller $X$, while the single-shock case dominates at larger $X$ and converges to $f_{\rm ml}=6\times10^{-4}$ for $X\gtrsim10^2$. This convergence reflects the fact that a star moving in a homogeneous medium will experience a mass loss proportional to the distance traveled  ($\Delta M_*\propto t \propto H$; see Eq.~\ref{eq:mass-loss-general}).

\begin{figure}
\centering
\includegraphics[width=\columnwidth]{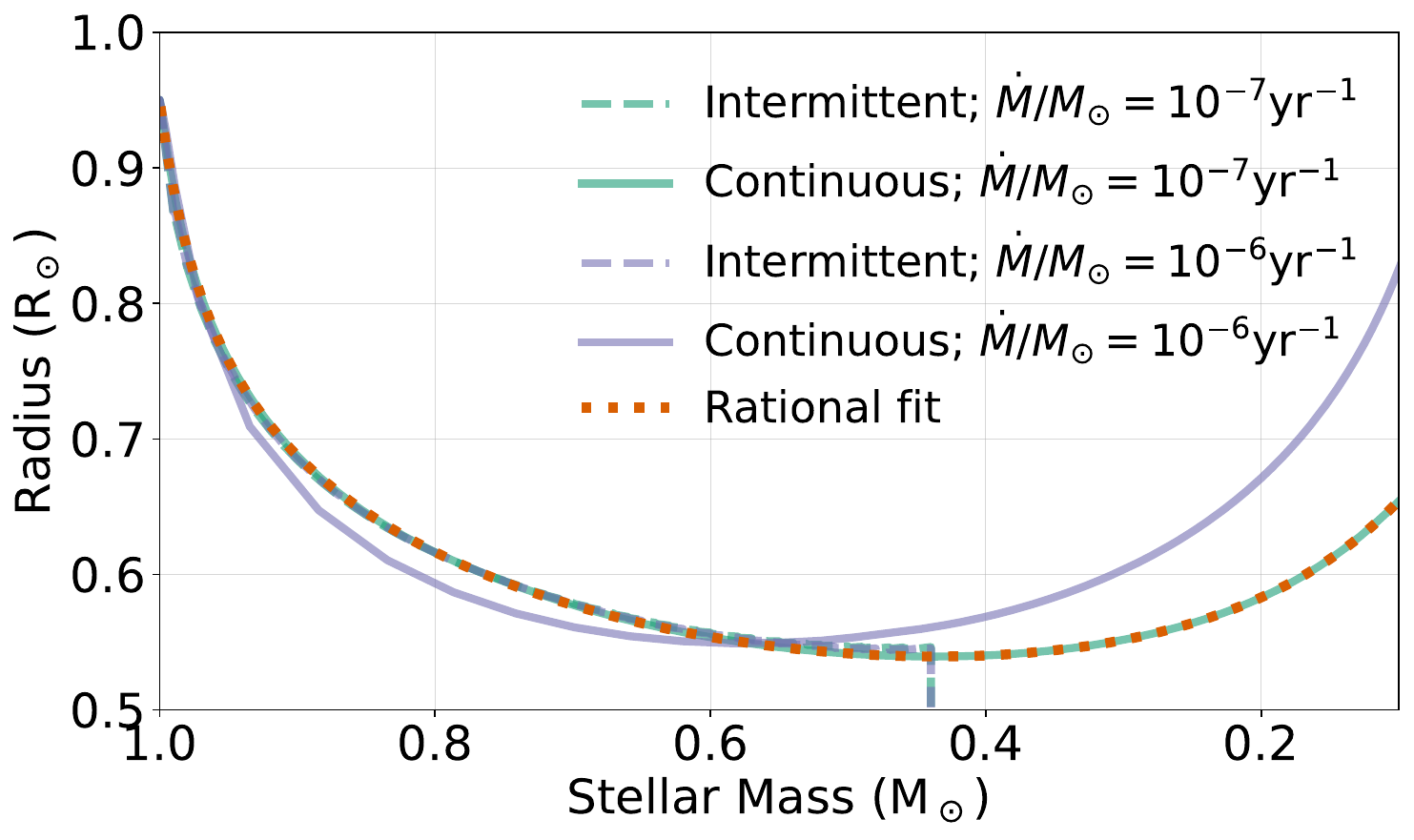}
\includegraphics[width=\columnwidth]{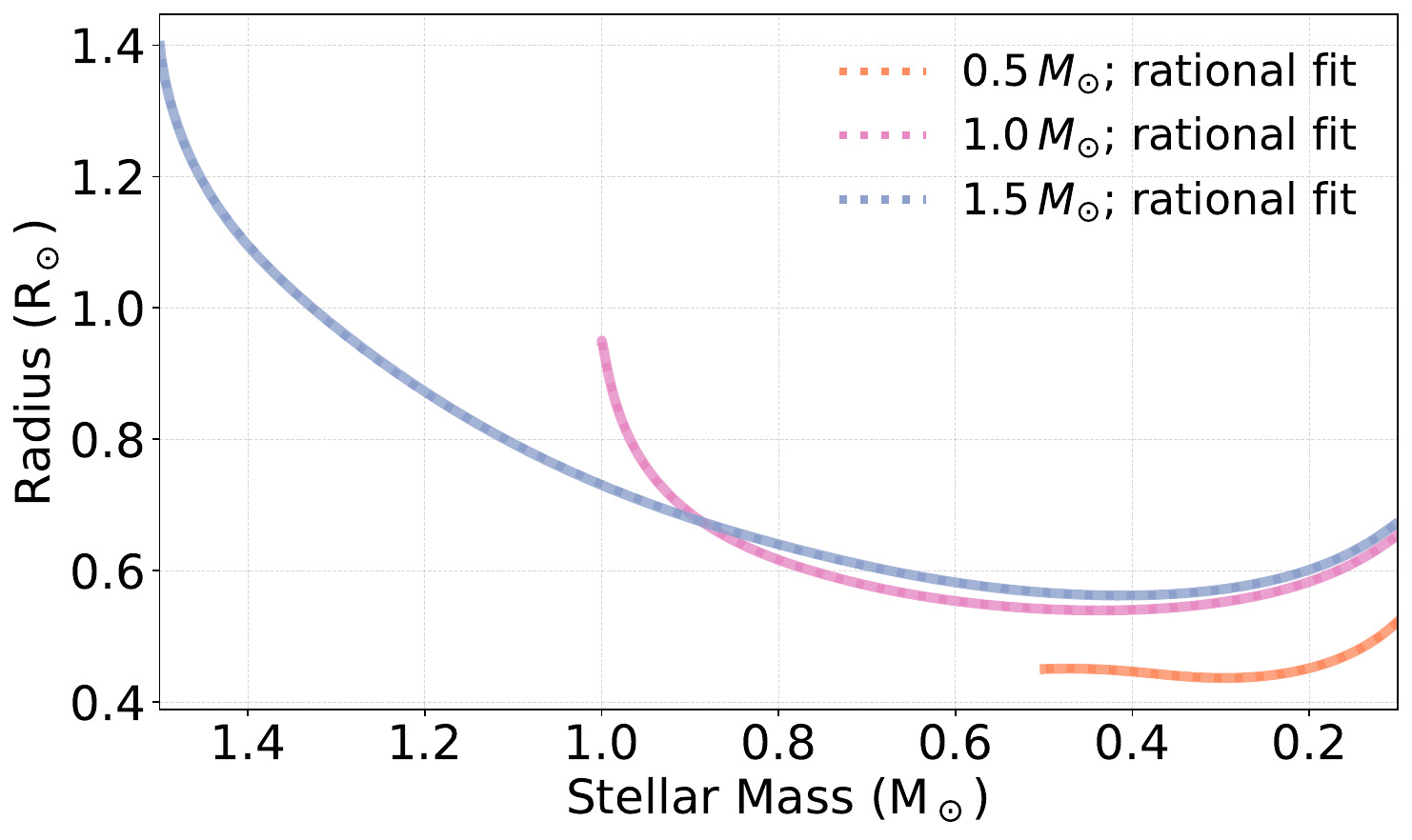}\\
\caption{Stellar radius as a function of mass for stars undergoing mass loss, computed with MESA. In both panels the star evolves from left to right as it loses mass. \emph{Top:} a solar-type ($1\,M_\odot$) star for two mass-loss rates, $\dot{M} = 10^{-7}$ (green) and $10^{-6}\,M_\odot\,\mathrm{yr}^{-1}$ (purple); solid curves assume continuous mass loss and dashed curves an intermittent prescription. The dotted orange curve is the rational fit (Eq.~\ref{eq:RM-fit}) to the continuous $\dot{M}=10^{-7}\,M_\odot\,\mathrm{yr}^{-1}$ track, which we adopt as the mass--radius relation throughout this paper. \emph{Bottom:} continuous mass loss at $\dot{M} = 10^{-7}\,M_\odot\, \mathrm{yr}^{-1}$ for three initial masses, $0.5$, $1.0$, and $1.5\,M_\odot$ (solid curves); dotted curves of the same color show the corresponding rational fits (Eq.~\ref{eq:RM-fit}), each anchored at the star's zero-age main-sequence radius; the fitted coefficients are listed in Table~\ref{tab:RM-fit}.}
\label{fig:mesa-RM}
\end{figure}

Mass loss also inevitably alters the stellar structure and changes its radius. Capturing the evolution of the stellar radius is crucial, as it determines the star's cross section and internal pressure, both of which directly affect the efficiency of gas drag and mass loss during star--disk collisions.
We adopt a mass-radius relation taken directly from a \texttt{MESA}-evolved zero-age main-sequence (ZAMS) star with initial mass \(1\,M_\odot\)\citep{Paxton2011,Paxton2013,Paxton2015,Paxton2018,Paxton2019}. In Figure \ref{fig:mesa-RM}, we test continuous stripping of the star at \(\dot{M} = 10^{-6},\,10^{-7}\,M_\odot\,\mathrm{yr}^{-1}\) and intermittent stripping in which the star loses mass at \(\dot{M} = 10^{-2}\,M_\odot\,\mathrm{yr}^{-1}\) for a single timestep every \(10^{4},\,10^{5}\,\mathrm{yr}\), respectively (resulting in the same time-averaged \(\langle \dot{M} \rangle\)). In both cases, the star initially contracts as mass is removed from its predominantly radiative envelope, but once it is stripped down to its nearly isentropic core, the mass--radius relation inverts and the radius grows with further mass loss, because mass is removed faster than the star can thermally readjust ($M_*/\dot M_*\lesssim t_{\rm KH}$). The second part can only be evolved for the continuous mass loss case as MESA crashes for unstable mass transfer at high mass transfer rates of \(\dot{M} = 10^{-2}\,M_\odot\,\mathrm{yr}^{-1}\). Because the stellar structure is largely determined by the cumulative mass removed over a thermal timescale, rather than by the instantaneous rate during each burst, the continuous and intermittent prescriptions produce nearly identical mass-radius relations; as a result, we only use the mass-radius relation fitted for the continuous \(\dot{M} = 10^{-7}\,M_\odot\,\mathrm{yr}^{-1}\) case for the remainder of the paper. Similarly, as shown in the bottom panel of Figure \ref{fig:mesa-RM}, we carry out the same exercise for a \(0.5\,M_\odot\) ZAMS star and a \(1.5\,M_\odot\) ZAMS star to test the effect of stellar properties on long-term orbital evolution as they differ in size, compactness, and entropy profile. 

The non-monotonic shape of these tracks, with initial contraction followed by re-expansion once stripping reaches the nearly isentropic core, cannot be captured accurately by a low-order polynomial. We therefore fit each track with a rational function of the fractional mass lost, $\xi\equiv1-M_*/M_{*,0}$:
\begin{equation}\label{eq:RM-fit}
    \frac{R_*}{R_{*,0}}=\frac{1+c_1\xi+c_2\xi^2+c_3\xi^3}{1+d_1\xi+d_2\xi^2+d_3\xi^3},
\end{equation}
where $M_{*,0}$ and $R_{*,0}$ are the ZAMS mass and radius of the \texttt{MESA} model. This form guarantees $R_*=R_{*,0}$ at $M_*=M_{*,0}$ for any choice of the coefficients, so every fit starts from the unstripped star. We determine the six coefficients by least squares over $0.1\,M_\odot\leq M_*\leq M_{*,0}$ and verify that the denominator remains positive over this range, so the fit has no poles. The fits reproduce the \texttt{MESA} radii to within $0.3\%$ everywhere (rms $\lesssim0.1\%$), whereas a quadratic anchored at the same point deviates by up to $16\%$. The fitted coefficients are listed in Table~\ref{tab:RM-fit}.

\begin{table*}
\centering
\caption{Coefficients of the rational mass--radius fit (Eq.~\ref{eq:RM-fit}) for continuous mass loss at $\dot{M}=10^{-7}\,M_\odot\,\mathrm{yr}^{-1}$, valid for $0.1\,M_\odot\leq M_*\leq M_{*,0}$.}
\label{tab:RM-fit}
\begin{tabular}{cccccccc}
\hline
$M_{*,0}$ $(M_\odot)$ & $R_{*,0}$ $(R_\odot)$ & $c_1$ & $c_2$ & $c_3$ & $d_1$ & $d_2$ & $d_3$\\
\hline
0.4990 & 0.4502 & $-2.591$ & $5.355$ & $-3.193$ & $-2.647$ & $5.961$ & $-4.057$\\
1.0000 & 0.9503 & $17.262$ & $-3.903$ & $-6.638$ & $25.696$ & $17.042$ & $-35.031$\\
1.4995 & 1.3989 & $55.322$ & $-57.766$ & $14.529$ & $66.303$ & $61.044$ & $-106.615$\\
\hline
\end{tabular}
\end{table*}

\section{Minimum Period From Disk Collisions Only} \label{sec:disk-only}
In this section, we model the long-term evolution of the stellar orbit and its properties under the influence of gas drag and mass loss of star--disk collisions. We isolate the disk effects and do not include GW orbital decay, which is explored in \S\ref{sec:disk-GW-model}.

Consider a star of mass $M_*$ in an orbit $(a,r_p)$ around the MBH of mass $\MBH$. The star is in repeated collisions with the time-averaged TDE disk characterized by $(\MBH,h,\Gamma_{\rm TDE})$. We can express the evolution of $(a,r_p,M_*)$ as functions of the number of orbits $N$:
\begin{align}\label{eq:star-evolution-diskonly}
    \frac{da}{dN} & = (\Delta a)_{\rm disk} = (\Delta a)_{\rm drag,peri} + (\Delta a)_{\rm drag,apo}, \nonumber\\
    \frac{dr_p}{dN} & = (\Delta r_p)_{\rm disk} = 2(\Delta a)_{\rm drag,apo}, \nonumber\\
    \frac{dM_*}{dN} & = (\Delta M_*)_{\rm disk} = (\Delta M_*)_{\rm peri} + (\Delta M_*)_{\rm apo},
\end{align}
where $(\Delta a)_{\rm disk}, (\Delta r_p)_{\rm disk}$, and $(\Delta M_*)_{\rm disk}$ are the per-orbit changes in $a,r_p,$ and $M_*$ due to star--disk collisions, respectively.
The per-orbit changes sum up the contributions from pericenter and apocenter, as indicated by subscripts $_{\rm peri}$ and $_{\rm apo}$.
Their expressions are given in Eq.~\ref{eq:a-drag}, \ref{eq:rp-drag}, and \ref{eq:mass-loss-general}.
The effects of collisions at the apocenter are only added when $r_a<r_d(t_f)$.
To solve Eq.~\ref{eq:star-evolution-diskonly}, we supplement it with the mass-radius relation $R_*(M_*)$ fitted from the \texttt{MESA} model (see \S\ref{sec:mass-loss}).

The above equations can be further simplified, giving the evolution of $r_p$ and $M_*$ as functions of $a$:
\begin{align}\label{eq:star-evolution-diskonly-simp}
    \frac{dr_p}{da} & = \frac{(dr_p/dN)}{(da/dN)} = \frac{(\Delta r_p)_{\rm disk}}{(\Delta a)_{\rm disk}}, \nonumber\\
    \frac{dM_*}{da} & = \frac{(dM_*/dN)}{(da/dN)} = \frac{(\Delta M_*)_{\rm disk}}{(\Delta a)_{\rm disk}}.
\end{align}
In our simulations, we set the initial stellar mass to $M_{*,0}=1\,M_\odot$.
For an initial orbit $(a_0,r_{p,0})$, Eq.~\ref{eq:star-evolution-diskonly-simp} is evolved until either condition below is satisfied:
\begin{enumerate}
    \item $M_*<M_{*,\min}=0.4\,M_\odot$: The star has experienced significant mass loss and is likely to start inflating (Figure~\ref{fig:mesa-RM}).
    \item $r_p<r_{p,\min}=2\,r_{t,\odot}$: Mass loss due to tidal stripping becomes likely.
\end{enumerate}
Our choice of $r_{p,\min}$ is conservative and is based on the initial condition of the star. In principle, the disruption radius is further affected by the stellar response to mass loss \citep[Figure~\ref{fig:mesa-RM}; see also][]{Bandopadhyay2025} and potential tidal heating \citep{Yao2026}. The more detailed modeling of the long term stellar response is beyond the scope of this work.

\begin{figure}
    \centering
    \includegraphics[width=\linewidth]{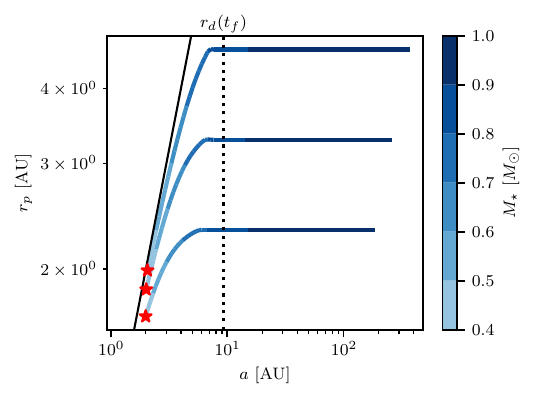}\\
    \caption{Examples of stellar mass and orbit evolution due to gas drag and mass loss of star--disk collisions (Eq.~\ref{eq:star-evolution-diskonly-simp}). The MBH mass is $\MBH=10^6\,M_\odot$, and the disk parameters are $h=0.1$ and $\Gamma_{\rm TDE}=10^{-4}\,{\rm yr}^{-1}$ (see \S\ref{sec:disk-model}). The star has initial mass $M_{*,0}=1\,M_\odot$ and eccentricity $e_0=e_{\rm Hills}$, typical of the Hills breakup (Eq.~\ref{eq:Hills-eccentricity}). The vertical line marks the maximum extent of the disk $r_d(t_f)$, and the diagonal line highlights the circular orbits. The endpoints of orbits, where $M_*=M_{*,\min}$, are marked with the red stars.}
    \label{fig:mass-orbit-evolution-diskonly}
\end{figure}

Figure~\ref{fig:mass-orbit-evolution-diskonly} shows several examples of stellar evolution under star--disk collisions. The initial orbits are chosen to simulate the conditions after the Hills breakup, with initial eccentricity
\begin{equation}\label{eq:Hills-eccentricity}
    e_0=e_{\rm Hills} = 1-\lrb{\frac{2\,M_\odot}{\MBH}}^{1/3}.
\end{equation}
In the beginning, the evolution of highly eccentric orbits is dominated by much more violent collisions at the pericenter, which only shrinks $a$ and not $r_p$ (Eq.~\ref{eq:rp-drag}).
Later, when the eccentricity drops and the orbit becomes more circular, collisions at the apocenter also start to contribute to gas drag and mass loss, and $r_p$ starts to shrink as well.
Eventually, the star loses a significant amount of mass and reaches $M_*=M_{*,\min}$.

More insight into the dynamics of the system can be gained by considering a simplified situation where the star only collides with the disk at the pericenter, which in most cases dominates the gas drag and mass loss. In this setup, $r_p$ is fixed and $M_*$ evolves as
\begin{equation}
    \frac{d\ln M_*}{d\ln a} = \frac{(\Delta M_*)_{\rm peri}/M_*}{(\Delta a)_{\rm drag,peri}/a}.
\end{equation}
We further replace $(\Delta M_*)_{\rm peri}$ and $(\Delta a)_{\rm drag,peri}$ with Eq.~\ref{eq:mass-loss-general} and \ref{eq:a-drag}, and expand the expressions of $p_{\rm ram},\,p_*,\,v_{\rm col},$ and $E$. This leads to the following equation:
\begin{equation}
    \frac{d\ln M_*}{d\ln a}=2\sqrt{\frac{\ln2}{\pi}}\frac{\MBH}{a}\frac{R_*}{M_*}f_{\rm ml},
\end{equation}
where $f_{\rm ml}=f_{\rm ml}(X=H(r_p)/R_*=h r_p/R_*)$.
Several observations can immediately be made from this equation:
\begin{enumerate}
    \item The majority of the stellar mass is lost at the smallest $a$, where the large magnitude of the orbital energy prevents the gas drag from further shrinking the orbit.
    \item The evolution depends on TDE disk properties only through $f_{\rm ml}$ and $h$, and is independent of $\lra{\Sigma}$ and $\Gamma_{\rm TDE}$. This results from the fact that both the mass loss and gas drag are linearly proportional to $p_{\rm ram}$ and $\lra{\Sigma}$ for a fixed disk height $H$.
    \item The stellar mass is depleted when $(d\ln M_*)/(d\ln a)$ increases to order unity and the mass loss becomes significant. The existence of a minimum in $f_{\rm ml}$ means that this inevitably happens at sufficiently small $a$, and suggests a minimum period $P_{\min}$ barrier for orbits that evolve solely due to star--disk collisions.
    \item Ignoring the variations in $M_*,R_*$, and $f_{\rm ml}$, $(d\ln M_*)/(d\ln a)\sim 1$ translates into a scaling in the minimum SMA and period:\begin{equation}\label{eq:P-min-scaling}
        a_{\min}\propto \MBH,\quad P_{\min}\propto \MBH.
    \end{equation}
    \item TDE disks with larger $h$ yield smaller $f_{\rm ml}$ and, consequently, shorter $P_{\min}$. This trend continues until $f_{\rm ml}$ reaches its minimum value of $\sim6\times10^{-4}$, beyond which $P_{\min}$ converges.
\end{enumerate}

\begin{figure}
    \centering
    \includegraphics[width=\linewidth]{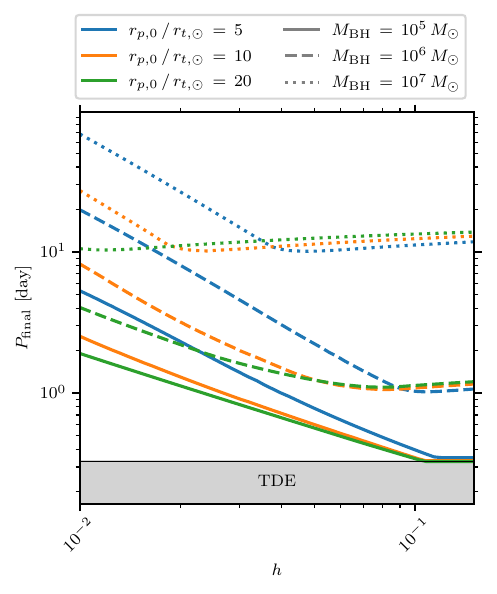}\\
    \caption{The final period $P_{\rm final}$ as a function of $h$ for different $r_{p,0}$ (color) and $\MBH$ (line style). Here we only consider the circularization from the gas drag and the mass loss from the disk collisions (Eq.~\ref{eq:star-evolution-diskonly-simp}). The initial orbits follow $e_0=e_{\rm Hills}$ (Eq.~\ref{eq:Hills-eccentricity}).  The TDE rate does not affect this result, but sets the time to reach the final orbital period.}
    \label{fig:P_final-h-diskonly}
\end{figure}

Figure~\ref{fig:P_final-h-diskonly} shows the final period $P_{\rm final}$ as a function of $h$ for different $r_{p,0}$ and $\MBH$. All orbits are started from the Hills breakup line $e_{\rm Hills}$ (Eq.~\ref{eq:Hills-eccentricity}). For a fixed pair of $\MBH$ and $r_{p,0}$, increasing $h$ initially leads to lower values of $P_{\rm final}$ due to the larger disk scale height at the point of collision ($H\propto h$) and decreased $f_{\rm ml}$. This trend continues until $f_{\rm ml}$ converges at a sufficiently large $h$, and $P_{\rm final}$ hits a lower limit. This lower limit is mainly set by $\min(f_{\rm ml})\approx6\times10^{-4}$ and is unaffected by $h$. Therefore, in the absence of other circularization mechanisms, there exists a period barrier that is independent of disk properties ($h$ and $\Gamma_{\rm TDE}$) and is set by the intrinsic mass loss physics of star--disk collisions.

\begin{figure}
    \centering
    \includegraphics[width=\linewidth]{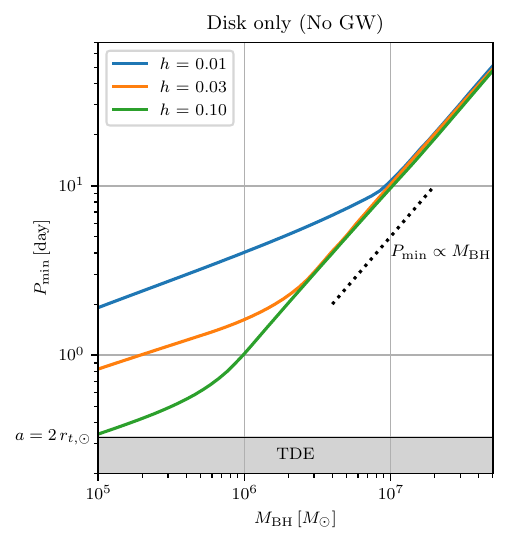}\\
    \caption{The minimum achievable period $P_{\min}$ of the Hills-captured star due to the drag and mass loss from star--disk collisions. For a given $\MBH$ and $h$, we compute $P_{\rm final}$ for orbits with $r_{p,0}/r_{p,\min} \in (1,10]$ and $e_0=e_{\rm Hills}$ (Eq.~\ref{eq:Hills-eccentricity}), and set $P_{\min}=\min(P_{\rm final})$. The TDE rate does not affect this result, but sets the time to reach the final orbital period.}
    \label{fig:P-min-diskonly}
\end{figure}

We next explore the minimum period $P_{\min}$ of Hills breakup stars under star--disk collisions. Formally, for a given set of $\MBH$, $h$ and $\Gamma_{\rm TDE}$, we define $P_{\min}$ as the minimum of $P_{\rm final}$ among the orbits initialized with $e_0=e_{\rm Hills}$ (Eq.~\ref{eq:Hills-eccentricity}).
Figure~\ref{fig:P-min-diskonly} shows $P_{\min}$ as a function of $\MBH$ for different values of $h$.
For large $\MBH\gtrsim10^7\,M_\odot$, $P_{\min}$ converges across different values of $h$ and shows the expected trend $P_{\min}\propto\MBH$ (Eq.~\ref{eq:P-min-scaling}).
This convergence reflects the period barrier and arises because the TDE disks at large $\MBH$ have sufficiently large $H$ that $f_{\rm ml}\approx6\times10^{-4}$ becomes nearly insensitive to $h$.
For lower $\MBH\lesssim 10^6\,M_\odot$, the disks with small $h$ yield values of $P_{\min}$ above the $P_{\min}\propto\MBH$ scaling because their smaller $H$ leads to larger $f_{\rm ml}$.

The trend of increasing $P_{\min}$ with $\MBH$ in Figure~\ref{fig:P-min-diskonly} highlights a major issue with the orbital circularization through disk drag alone. In the absence of other ways to circularize the orbit more efficiently, the collisions with the disk will destroy the star before it can reach sub-day periods for $\MBH\gtrsim 10^6\,M_\odot$.
This period barrier is set by the mass loss physics during star--disk collisions and cannot be reduced with a thicker disk due to the minimum value of $f_{\rm ml}$.
The inability to produce orbits of sub-day periods is in tension with the stellar orbiter models for some observed QPE sources.
In the next section, we explore the additional effect of GW orbital decay and attempt to alleviate the tension at large $\MBH$.

\section{Minimum Period From Disk Collisions \& Gravitational Wave}\label{sec:disk-GW-model}
GW emission provides an additional mechanism for orbital circularization. If GW-driven orbital decay becomes comparable to or more efficient than the gas drag produced by star--disk collisions, the stellar orbit can overcome the $P_{\min}$ barrier imposed by star--disk collisions. In this section, we incorporate GW orbital decay into our stellar evolution model and investigate the conditions under which it significantly alters the star's final orbit.

\subsection{GW orbital decay versus TDE gas drag}
The GW orbital decay is purely dynamical and does not cause additional mass loss to the star.
The per-orbit change due to GW emission is \citep{Peters1964}:
\begin{align}
    (\Delta a)_{\rm GW}= & -\frac{128\pi}{5}a\lrb{\frac{a}{r_{g}}}^{-5/2}\lrb{\frac{M_{*}}{M_{\text{BH}}}}\nonumber\\
    & \times\frac{1}{(1-e^{2})^{7/2}}\lrb{1+\frac{73}{24}e^{2}+\frac{37}{96}e^{4}},\\
    (\Delta r_p)_{\rm GW}=& -\frac{2\pi}{15}a\lrb{\frac{a}{r_{g}}}^{-5/2}\lrb{\frac{M_{*}}{M_{\text{BH}}}}\nonumber\\
    &\times\frac{(1-e)^{2}}{(1-e^{2})^{7/2}}\lrb{47e^{3}+168e^{2}-112e+192}.
\end{align}
Combined with the effects from star--disk collisions, the orbit evolution in Eq.~(\ref{eq:star-evolution-diskonly}) is modified and becomes
\begin{align}\label{eq:star-evolution-GWdisk-full}
    \frac{da}{dN} & = (\Delta a)_{\rm disk}+(\Delta a)_{\rm GW},\nonumber\\
    \frac{dr_p}{dN} & = (\Delta r_p)_{\rm disk} + (\Delta r_p)_{\rm GW}.
\end{align}
The mass loss evolution stays unchanged. As in \S\ref{sec:disk-only}, the above equations can be simplified by writing $r_p$ and $M_*$ as functions of $a$:
\begin{align}\label{eq:star-evolution-GWdisk-simp}
    \frac{dr_p}{da} & = \frac{(\Delta r_p)_{\rm disk}+(\Delta r_p)_{\rm GW}}{(\Delta a)_{\rm disk}+(\Delta a)_{\rm GW}}, \nonumber\\
    \frac{dM_*}{da} & = \frac{(\Delta M_*)_{\rm disk}}{(\Delta a)_{\rm disk}+(\Delta a)_{\rm GW}}.
\end{align}
The stopping conditions are the same as in \S\ref{sec:disk-only}.

\subsection{Circularization timescales}\label{sec:circularization-timescales}

\begin{figure}
    \centering
    \includegraphics[width=\linewidth]{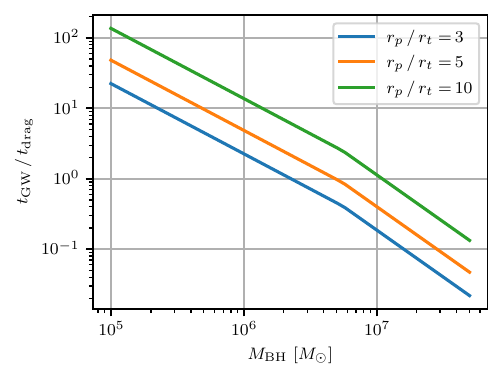}\\
    \caption{The timescale ratio between GW orbit decay (Eq.~\ref{eq:t_GW}) and the gas drag circularization (Eq.~\ref{eq:t_drag}) as a function of $\MBH$. The disk parameters are fixed to $h=0.1$ and $\Gamma_{\rm TDE}=10^{-4}\,{\rm yr}^{-1}$. The star has $M_*=1\,M_\odot$ and $R_*=1\,R_\odot$. We show result for three orbits, with $r_p=3,5,10\,r_{t,\odot}$. The eccentricities of the orbits are given by Eq.~(\ref{eq:Hills-eccentricity}), although the timescale ratio is insensitive to $a$ and only depends on $r_p$ (Eq.~\ref{eq:t_drag-t_GW-ratio}).  For lower mass BHs, disk drag is more important than GW orbital decay in circularizing EMRI orbits.}
    \label{fig:t_drag-t_GW-ratio}
\end{figure}

\begin{figure*}
    \centering
    \includegraphics[width=\textwidth]{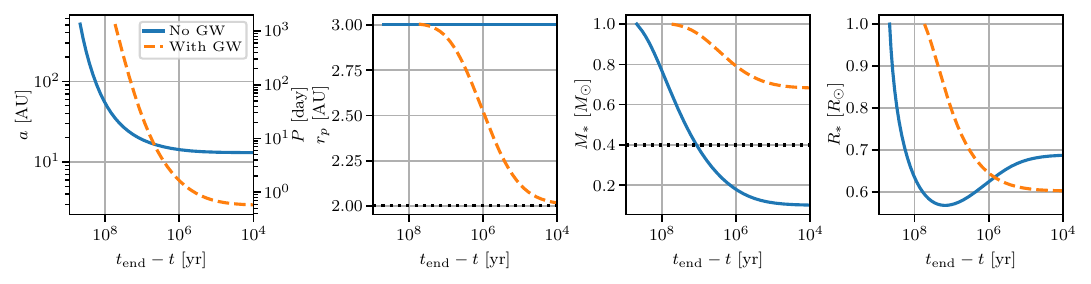}\\
    \caption{Time evolution of the stellar and orbital properties. From left to right, each panel shows SMA $a$ and period $P$, pericenter radius $r_p$, stellar mass $M_*$, and stellar radius $R_*$.
    The blue curves show the evolution with only disk effects (gas drag and mass loss), while the orange curves also include GW orbital decay. The MBH mass is $\MBH=10^7\,M_\odot$, and the TDE disk parameters are $h=0.1$ and $\Gamma_{\rm TDE}=10^{-4}\,{\rm yr}^{-1}$. The star initially has $r_p=3\,r_{t,\odot}$ and eccentricity given by Eq.~(\ref{eq:Hills-eccentricity}). The horizontal axis shows the time remaining until the end of the simulation, ($t_{\rm end}-t$). We omit the interval ($t_{\rm end}-t<10^4\,{\rm yr}$), where the time-averaged treatment of the TDE disk is expected to become inaccurate. The black dotted line in the pericenter radius panel shows $r_{p,\min}=2\,r_{t,\odot}$. For this figure, we lower the minimum stellar mass to $M_{*,\min}=0.1\,M_\odot$ to illustrate the possible evolution after the star begins to inflate. In all other calculations, we adopt the more conservative value $M_{*,\min}=0.4\,M_\odot$ shown by the black dotted line.
    \label{fig:mass_loss_evolution-GWdisk}}
\end{figure*}

Before solving the system numerically, it is instructive to examine analytically how GW emission affects the orbital evolution. The GW orbital decay can only help the star reach significantly shorter-period orbits if its circularization timescale is comparable to or smaller than that of the gas drag from star--disk collisions.
For gas drag, the circularization timescale is
\begin{equation}\label{eq:t_drag}
    t_{\rm drag}=P\cdot\abs{\frac{a}{(\Delta a)_{\rm drag}}}.
\end{equation}
For highly eccentric orbits, gas drag is dominated by collisions at the pericenter, giving
\begin{align}
    t_{\rm drag} & \sim \abs{\frac{a}{(\Delta a)_{\rm drag,peri}}} P\nonumber\\
    & \sim \lrsb{\frac{M_{*}}{\pi R_{*}^{2}\cdot\lra{\Sigma}(r=r_{p})}}\lrb{\frac{r_{p}}{a}}P \nonumber\\
    & \propto \MBH^{-1/6}r_p^2a^{1/2}.
\end{align}
In the last line we assume $r_{p,\min}<r_p<r_d(t_f)$ and use the following scaling:
\begin{equation}
    \lra{\Sigma}(r_p)\propto\Sigma_{d,0}\lrb{\frac{r_{d,0}}{r_p}}\propto\MBH^{-1/3}r_p^{-1}.
\end{equation}
For clarity, we omit the dependence on the TDE disk properties $(M_{d,0},h,\Gamma_{\rm TDE})$ and the stellar properties $(M_*,R_*)$.
For GW emission, the circularization timescale of highly eccentric orbits is
\begin{align}\label{eq:t_GW}
    t_{\rm GW} &=P\cdot\abs{\frac{a}{(\Delta a)_{\rm GW}}}\nonumber\\
    & \sim \frac{12\sqrt{2}}{85}\lrb{\frac{r_g}{c}}\lrb{\frac{M_{\rm BH}}{M_*}}\lrb{\frac{a}{r_g}}^{1/2}\lrb{\frac{r_p}{r_g}}^{7/2}\nonumber\\
    & \propto \MBH^{-2}r_p^{7/2}a^{1/2}.
\end{align}
The ratio between the two timescales scales with the orbital parameters as
\begin{equation}\label{eq:t_drag-t_GW-ratio}
    \frac{t_{\rm GW}}{t_{\rm drag}}\propto \MBH^{-11/6}r_p^{3/2}a^{0}.
\end{equation}
Since this ratio is independent of $a$, the dominant circularization mechanism for a highly eccentric orbit is determined solely by $r_p$.
For the Hills-captured stars with $r_p\propto (a_b/R_*)r_{t,\odot}\propto \MBH^{1/3}$, the scaling is further simplified into
\begin{equation}
\frac{t_{\rm GW}}{t_{\rm drag}}\propto\MBH^{-4/3}.
\end{equation}
Therefore, GW-driven orbital circularization becomes increasingly important around more massive black holes.

Figure~\ref{fig:t_drag-t_GW-ratio} shows $t_{\rm GW}/t_{\rm drag}$ as a function of $\MBH$ for different $r_p$. The numerical results agree with the analytical calculation that the ratio decreases with increasing BH masses. The sudden change in slope near $\MBH\sim10^7\,M_\odot$ is due to the dependence of $M_{d,0}$ on $\MBH$, which is not captured in the analytical scalings.
In general, the GW orbital decay outpaces the gas drag for $\MBH\gtrsim10^{6-7}\,M_\odot$, with the exact values depending on the stellar properties, pericenter radius $r_p$, and TDE disk properties ($M_{d,0},\Gamma_{\rm TDE},h$).

% \begin{figure}
%     \centering
%     \includegraphics[width=\linewidth]{Figures/mass_loss_evolution_MBH7.0_h0.1.pdf}\\
%     \caption{The evolution of stellar mass $M_*$ as a function of SMA $a$ with the inclusion of GW effect. The dot-dashed lines show the result with only disk drag (Eq.~\ref{eq:star-evolution-diskonly-simp}). The solid lines show the result where the GW orbital decay is included (Eq.~\ref{eq:star-evolution-GWdisk-simp}). The BH mass and TDE disk properties are the same as in Figure~\ref{fig:mass-orbit-evolution-diskonly} \howard{The caption is outdated - this figure uses $\MBH=10^7\,M_\odot$}. The stellar orbits have initial SMA set by the eccentricity after the Hills breakup (Eq.~\ref{eq:Hills-eccentricity}).}
%     \label{fig:mass_loss_evolution-GWdisk-old}
% \end{figure}

\subsection{GW effect on minimum periods}

The acceleration of circularization from the GW emission allows the star to break the period barrier set by star--disk collisions.
Figure~\ref{fig:mass_loss_evolution-GWdisk} compares two stellar evolutions with identical initial conditions, one including GW orbital decay and the other neglecting it. At the high BH mass of $\MBH=10^7\,M_\odot$, the star is completely depleted at $P_{\rm final}\sim10\,{\rm days}$ without GW orbital decay ($P_{\rm final}\propto\MBH$; see \S\ref{sec:disk-only}). In contrast, the strong GW emission ($t_{\rm GW}<t_{\rm drag}$) enables the star to reach $P_{\rm final}<1\,{\rm day}$, where it is ultimately destroyed upon reaching $r_{p,\min}=2\,r_{t,\odot}$.

\begin{figure}
    \centering
    \includegraphics[width=\linewidth]{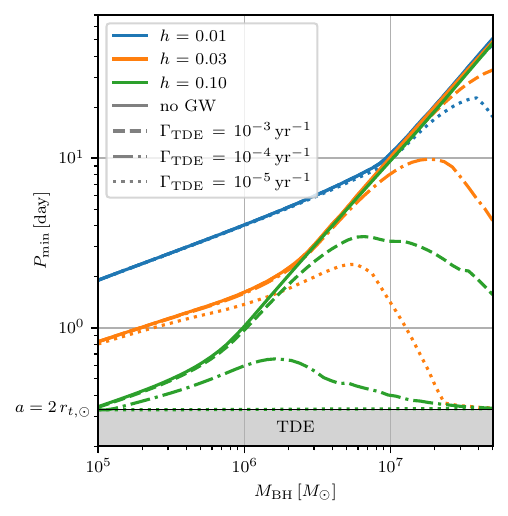}\\
    \caption{The minimum achievable period $P_{\min}$ of the Hills-captured star due to  GW orbital decay, disk drag and mass loss from collisions with TDE disks.}
    \label{fig:P-min-GWdisk}
\end{figure}

In Figure~\ref{fig:P-min-GWdisk}, we further quantify the GW effects by showing the minimum achievable period $P_{\min}$ as functions of $\MBH$ for a range of disk parameters $0.01\le h\le0.1$ (different colors) and $10^{-5}\,{\rm yr}^{-1}\le\Gamma_{\rm TDE}\le10^{-3}\,{\rm yr}^{-1}$ (different line styles). We solve the orbits with Eq.~\ref{eq:star-evolution-GWdisk-simp} and use the same orbit sampling method as in Figure~\ref{fig:P-min-diskonly}. The case without GW orbital decay is shown again here for comparison.
For $\MBH\lesssim10^6\,M_\odot$, the effect of GW is mild due to its relatively long timescale (Eq.~\ref{eq:t_drag-t_GW-ratio}), and is only noticeable for a thicker disk ($h=0.1$) and a lower TDE rate ($\Gamma_{\rm TDE}=10^{-4}\,{\rm yr}^{-1}$). For $\MBH\gtrsim10^6\,M_\odot$, the GW circularization becomes very efficient and is able to bring the star to much smaller orbits before the mass loss from star--disk collision or tidal disruption destroys the star.
With the inclusion of GW emission, stellar orbits at higher BH masses ($\MBH\gtrsim10^6\,M_\odot$) can also reach sub-day periods for $h=0.1$ and $\Gamma_{\rm TDE}\lesssim10^{-4}\,{\rm yr}^{-1}$. At very low TDE rates $\Gamma_{\rm TDE}\lesssim10^{-5}\,{\rm yr}^{-1}$, the orbits can be circularized until they reach the disruption limit $2\,r_{t,\odot}$.

\section{QPE Rate Suppression from Relaxation}\label{sec:full-model}
Our models in \S\ref{sec:disk-only} and \ref{sec:disk-GW-model} show that the stellar orbits produced by Hills breakup can circularize and reach sub-day periods through gas drag from star--disk collision at $\MBH\lesssim10^6\,M_\odot$ and through GW emission at $\MBH\gtrsim10^6\,M_\odot$.
However, a complete orbit evolution model must also consider the orbital relaxation from gravitational encounters.
Unlike gas drag and GW emission that preserve pericenter radii during orbit circularization, the relaxation perturbs the orbital angular momentum and can cause the pericenter to cross the tidal radius.
In this section, we develop a Monte Carlo model that incorporates the stochastic relaxation process, and use it to explore the interplay with star--disk collisions, GW emission, and relaxation.

\subsection{Full orbital evolution model including relaxation}
The Keplerian stellar orbits around the MBH in galactic nuclei are gradually changed by the random gravitational encounters with other stars/compact objects. In our model, we ignore the details of relaxation and use a uniform two-body relaxation timescale $t_{\rm 2B,rel}$ to quantify its strength. We only consider the relaxation of orbital angular momentum and ignore the relaxation of orbital energy, which is much slower for highly eccentric orbits produced by the Hills breakup.  We also do not consider resonant relaxation, as the high eccentricities of Hills-captured stars induce rapid relativistic apsidal precession that suppresses the coherent torques driving resonant relaxation \citep{Antonini2013,Generozov2020}.

For a stellar orbit with angular momentum $L$, its change after time $\Delta t$ is modeled as a Gaussian distribution:
\begin{equation}\label{eq:DeltaL-Gaussian}
    \frac{\Delta L}{L_c}\sim \mathcal{N}\lrb{\mu=\frac{L_c}{2L}\frac{\Delta t}{t_{\rm 2B,rel}},\sigma^2=\frac{\Delta t}{t_{\rm 2B,rel}}},
\end{equation}
where $L_c=L_c(a)=M_*\sqrt{GM_{\rm BH}a}$ is the angular momentum of the circular orbit with the same SMA.
The angular momentum relaxation timescale $t_{L,\rm rel}$ is the time that we expect a significant change in $L$ ($\abs{\Delta L}\sim L$):
\begin{equation}\label{eq:t-L-rel}
    t_{L,\rm rel} = \lrb{\frac{L^2}{L_c^2}}t_{\rm 2B,rel}=\lrb{\frac{2\,r_p}{a}}t_{\rm 2B,rel},
\end{equation}
where the second equal sign assumes a highly eccentric orbit.
The change in $L$ can be mapped to the change in $r_p$:
\begin{equation}\label{eq:Delta-rp-DeltaL}
\Delta r_p=a\lrsb{\sqrt{1-\lrb{\frac{L}{L_c}}^2}-\sqrt{1-\lrb{\frac{L+\Delta L}{L_c}}^2}}.
\end{equation}
This allows us to incorporate orbit relaxation into our orbit evolution model.

Due to the stochasticity of the relaxation process, orbital evolution cannot be solved by direct integration of differential equations. Instead, we modify Eq.~(\ref{eq:star-evolution-GWdisk-full}) and solve the orbit evolution iteratively, each time advancing the number of orbits by $\Delta N$:
\begin{align}
    a & \rightarrow a + \lrsb{(\Delta a)_{\rm disk}+(\Delta a)_{\rm GW}}\cdot\Delta N,\nonumber\\
    r_p & \rightarrow r_p + \lrsb{(\Delta r_p)_{\rm disk} + (\Delta r_p)_{\rm GW}}\cdot\Delta N+(\Delta r_p)_{\rm rel},\nonumber\\
    M_* & \rightarrow M_* + (\Delta M_*)_{\rm disk}\cdot\Delta N,\nonumber\\
    t & \rightarrow t+P\cdot \Delta N.
\end{align}
The amount of change in $r_p$ due to relaxation, $(\Delta r_p)_{\rm rel}$, is determined by randomly drawing $\Delta L$ from the Gaussian distribution in Eq.~(\ref{eq:DeltaL-Gaussian}) with $\Delta t=P\cdot \Delta N$, and then converting $\Delta L$ into $\Delta r_p$ using Eq.~(\ref{eq:Delta-rp-DeltaL}).
To minimize the number of iterations and reduce computation, $\Delta N$ is dynamically set at each iteration:
\begin{multline}
    \Delta N = \max \left[ \epsilon \cdot \min \left( \frac{a}{|(\Delta a)_{\rm disk} + (\Delta a)_{\rm GW}|}, \right. \right. \\
    \left. \left. \frac{M_*}{|(\Delta M_*)_{\rm disk}|}, \frac{t_{L\rm,rel}}{P} \right), 1 \right].
\end{multline}
This expression selects the step size based on the fastest process among orbital decay, mass loss, and angular momentum relaxation.
An additional dimensionless parameter $\epsilon$ is used to control the overall step size of iterations.
From our testing, $\epsilon=10^{-3}$ provides a good computation speed-up without compromising numerical accuracy. For a given initial orbit ($a_0,r_{p,0}$) and stellar mass $M_{*,0}=M_\odot$, we iteratively update the orbit until one of the following stopping conditions is reached:
\begin{enumerate}
    \item $r_{p}<r_{p,\min}=2\,r_{t,\odot}$
    \item $M_*<M_{*,\min}=0.4\,M_\odot$
    \item $t>t_{\max}=1\,{\rm Gyr}$
\end{enumerate}
The third stopping condition ensures that we exclude system outcomes that can only be reached with an excessively long time. The parameters used in the Monte Carlo simulations are summarized in Table~\ref{tab:model-parameters}.

\begin{table*}
	\centering
	\caption{Parameters adopted in our Monte Carlo orbit evolution model.}
	\label{tab:model-parameters}
	\begin{tabular}{lcl}
		\hline
		Symbol & Values/Ranges & Description\\
        \hline
        $M_{\rm BH}$ & $10^5-10^7\,M_\odot$ & MBH Mass\\
        $M_{*,0}$ & $M_{\odot}$ & Initial mass of the star\\
        $R_{*,0}$ & $R_{\odot}$ & Initial radius of the star\\
        $M_{*,\min}$ & $0.4\,M_\odot$ & Minimum mass of the star\\
        $r_{p,\min}$ & $2\,r_{t,\odot}$ & Minimum pericenter radius of the star\\
        $M_{d,0}$ & Eq.~(\ref{eq:M-d0}) & Initial mass of the TDE disk\\
        $\alpha$ & 0.1 & Viscosity parameter of the TDE disk\\
        $h$ & $0.01-0.1$ & TDE disk scale height to radius ratio\\
        $r_d(t_f)$ & $10\,r_{d,0}$ & Maximum extent of the TDE disk (see \S\ref{sec:disk-model})\\
        $\Gamma_{\rm TDE}$ & $0-10^{-3}\,{\rm yr}^{-1}$ & TDE rate per galaxy\\
        $t_{\rm 2B,rel}$ & $10^9-10^{11}\,{\rm yr}$ & Two-body relaxation time\\
		\hline
	\end{tabular}
    \vspace{0.6cm}
\end{table*}

\begin{figure*}
    \centering
    \includegraphics[width=\textwidth]{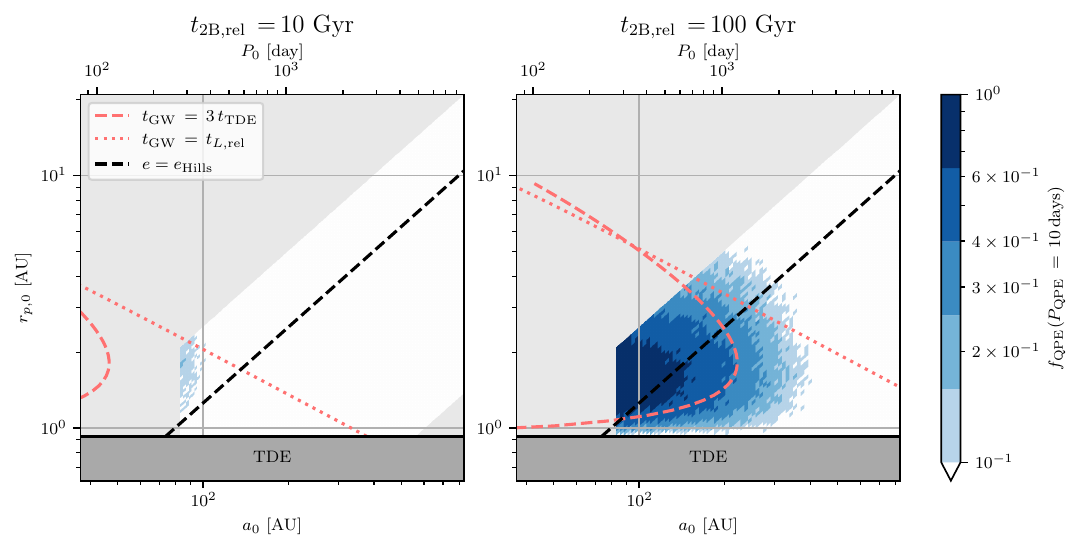}\\
    \caption{Fraction of stars that reach $P<10\,\rm days$ on a grid of initial orbits. The BH mass is $M_{\rm BH}=10^6\,M_\odot$ and we do not include star-disk collisions here.
    Each grid point represents 300 Monte Carlo samples. The left and right panels show results for $t_{\rm 2B,rel}=10$ and $100\,{\rm Gyr}$, respectively. The region not sampled in our simulation is covered in light gray, and the region of TDE ($r_p<r_{p,\min}$) is colored in dark gray. The conventional EMRI boundary $t_{\rm GW}=t_{L,\rm rel}$ and our proposed new boundary $t_{\rm GW}=3\,t_{\rm TDE}$ are plotted in red dotted and dashed lines. The typical eccentricity for the bound star after Hills breakup, $e_{\rm Hills}$, is highlighted as the black dashed line.
    \label{fig:QPE-fraction-grid-no-disk}}
\end{figure*}

\subsection{Orbit evolution without disk collisions}
We first explore the case without disk collisions to assess the ability of GW to form EMRIs and compete against the potential tidal disruption due to relaxation. We briefly review the conventional analytical argument that separates the GW-dominated and relaxation-dominated regimes, and use our Monte Carlo simulation to demonstrate how this argument severely over-predicts the EMRI rate due to the incomplete treatment of relaxation and tidal disruption.
\subsubsection{Conventional criterion of EMRI formation}
Conventionally, the separation between the GW-dominated and relaxation-dominated regimes is defined by equating the GW timescale $t_{\rm GW}$ (Eq.~\ref{eq:t_GW}) and the angular momentum relaxation timescale $t_{L,\rm rel}$ (Eq.~\ref{eq:t-L-rel}).
The resulting GW-relaxation boundary is \citep{Sari2019,linialSari2023}
\begin{equation}\label{eq:GW-rel-trad-bd}
    r_p = a_{\rm GW}\lrb{\frac{a}{a_{\rm GW}}}^{-3/5},
\end{equation}
where
\begin{align}
    a_{\rm GW}&=r_g\cdot\lrsb{\frac{85}{6\sqrt{2}}\lrb{\frac{M_*}{M_{\rm BH}}}\lrb{\frac{ct_{\rm 2B,rel}}{r_g}}}^{1/4}\nonumber\\
    &=5.0\,{\rm AU}\,\lrb{\frac{t_{\rm 2B,rel}}{\rm Gyr}}^{1/4}\lrb{\frac{M_{\rm BH}}{10^6\,M_{\odot}}}^{1/2}\lrb{\frac{M_*}{M_{\odot}}}^{1/4}.
\end{align}
For $r_p$ below this boundary, the orbit evolution is dominated by GW and stars can become EMRIs.
The fact that this boundary has decreasing $r_p$ with $a$ means that tidal disruption sets a maximum SMA $a_{\rm GW,\max}$ of the orbits that can be circularized by GW:
\begin{equation}
    a_{\rm GW,\max}=a_{\rm GW}\lrb{\frac{r_{p,\min}}{a_{\rm GW}}}^{-5/3}.
\end{equation}
Beyond $a_{\rm GW,\max}$, relaxation is dominant for all $r_p\gtrsim r_{p,\min}$, and stars are likely to become TDEs.
% In order for the Hills breakup stars to form EMRIs, the eccentricity of the GW-relaxation boundary (Eq.~\ref{eq:GW-rel-trad-bd}) must be greater than $e_{\rm Hills}$ (Eq.~\ref{eq:Hills-eccentricity}) at $r_p=r_{p,\min}$:
% \begin{equation}
%     \frac{r_{p,\min}}{a_{\rm GW,\max}}\lesssim1-e_{\rm Hills}.
% \end{equation}
% This constraint can be inverted to estimate the minimum two-body relaxation time for GW-EMRI formation:
% \begin{equation}\label{eq:t-rel-min-trad}
%     t_{\rm 2B,rel}\gtrsim9\times10^8\,{\rm yr}\lrb{\frac{\MBH}{10^6\,M_\odot}}^{-1/6}.
% \end{equation}

\subsubsection{Inefficiency of GW orbital decay}
To test the validity of the above conventional argument in predicting EMRI formation through GW orbital decay, we use our Monte Carlo model to simulate the outcomes of the stars. We temporarily set $\Gamma_{\rm TDE}=0\,{\rm yr}^{-1}$ in our model to exclude the effects of disk collisions. 
Figure~\ref{fig:QPE-fraction-grid-no-disk} shows the fraction of stars that avoid tidal disruption and circularize to periods less than 10 days on a grid of initial orbits.
The left and right panels correspond to $t_{\rm 2B,rel}=10$ and $100\,{\rm Gyr}$, respectively. The conventional GW-relaxation boundary (Eq.~\ref{eq:GW-rel-trad-bd}) is highlighted in red dotted lines.
Our Monte Carlo simulations show that a significant fraction of stars with initial orbits below Eq.~(\ref{eq:GW-rel-trad-bd}) fail to reach $P<10\,{\rm days}$, and that the conventional boundary severely overestimates the fraction of stars that can avoid tidal disruption and become EMRIs.

The discrepancy with the conventional argument can be understood by distinguishing the timescales for angular momentum relaxation and tidal disruption.
The angular momentum relaxation time, $t_{L,\rm rel}$, is the timescale over which order-unity changes in $L$ and $r_p$ are expected.
However, for a stellar orbit with $r_p\sim r_{p,\min}$, even a small change of $|\Delta r_p|\ll r_p$ can cause the pericenter to cross $r_{p,\min}$ and result in a TDE.
Therefore, the relevant timescale for stellar survival is not $t_{L,\rm rel}$ but rather $t_{\rm TDE}$, the timescale on which relaxation drives a star across the tidal-disruption boundary. We formalize the definition of $t_{\rm TDE}$ below.

Let $L_{\min}$ be the minimum angular momentum the star can have without being tidally disrupted:
\begin{equation}
    L_{\min}=M_*\sqrt{GM_{\rm BH}a\lrsb{1-\lrb{1-\frac{r_{p,\min}}{a}}^2}}.
\end{equation}
We define the local tidal disruption timescale $t_{\rm TDE}$ as the characteristic timescale on which orbital relaxation drives stars from a given initial orbit across the tidal-disruption boundary:
\begin{align}
t_{\rm TDE} = \frac{(L - L_{\min})^2}{L_c^2(a)} t_{\rm 2B,rel}.
\end{align}
For $r_p\sim r_{p,\min}$, we have $L-L_{\min}\ll L$ and therefore $t_{\rm TDE}\ll t_{L,\rm rel}$.
Figure~\ref{fig:QPE-fraction-grid-no-disk} shows that the following relation provides substantially better agreement with the EMRI--TDE boundary obtained from our Monte Carlo simulations:
\begin{equation}\label{eq:GW-rel-bd-new}
t_{\rm GW}=3\,t_{\rm TDE}.
\end{equation}
The curvature of this boundary toward smaller SMA at $r_p\lesssim2\,r_{p,\min}$ arises from the vanishingly short $t_{\rm TDE}$ near $r_p=r_{p,\min}$.
Nonetheless, because relaxation is intrinsically stochastic, the EMRI--TDE transition is considerably more blurry than suggested by this analytical criterion, and the relation $t_{\rm GW}=3\,t_{\rm TDE}$ should only be treated as an approximation of the boundary.

The tendency of the conventional criterion $t_{L,\rm rel}=t_{\rm GW}$ to overestimate the number of GW-driven EMRIs was also noted by \cite{Kaur2025}. They found that the bulk motion of the stellar population in the $(a,r_p)$ phase space only becomes strongly affected by GW at pericenter radii that are a fraction of those predicted by this criterion.

\begin{figure*}
    \centering
    \includegraphics[width=\textwidth]{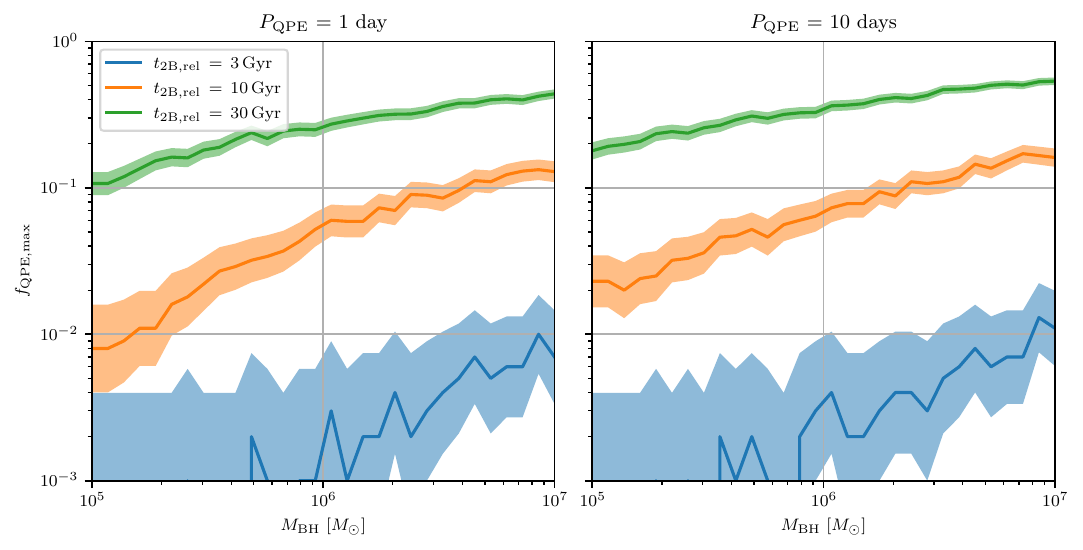}\\
    \caption{The maximum QPE fraction $f_{\rm QPE,\max}$ as a function of $\MBH$ for different $t_{\rm 2B,rel}$. We assume no disk collisions here.
    The shaded regions are the $2\sigma$ confidence intervals computed with \texttt{astropy.stats.binom\_conf\_interval} and \texttt{interval='wilson'}. The left and right panels correspond to QPE period limits of 1 and 10 days, respectively. \label{fig:f-QPE-max-no-disk}}
\end{figure*}

\begin{figure*}
    \centering
    \includegraphics[width=\textwidth]{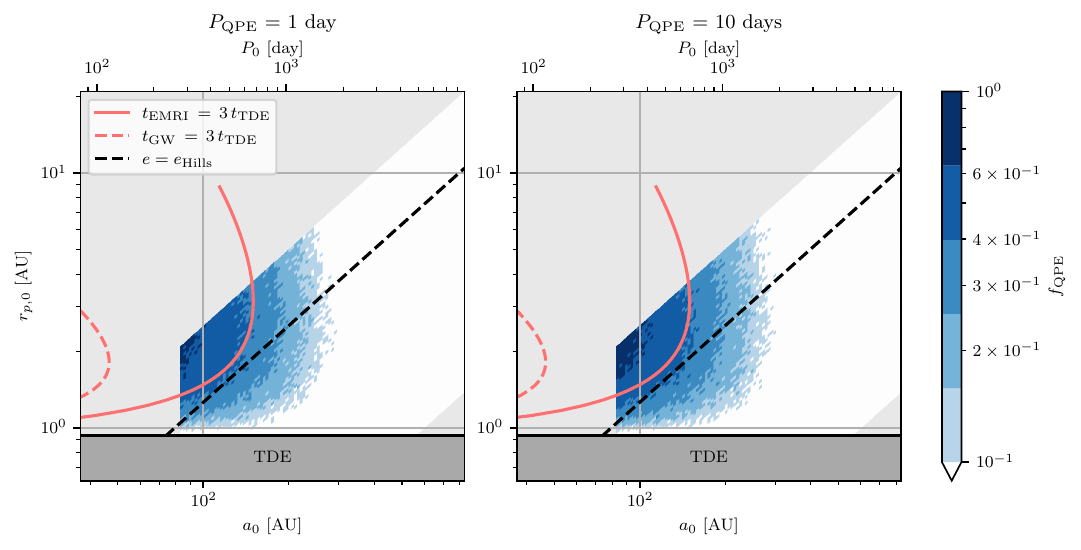}\\
    \caption{Fraction of (Hills-captured) stars that reach QPE orbits $f_{\rm QPE}$ on a grid of initial orbits for our fiducial model $(\MBH,h,\Gamma_{\rm TDE},t_{\rm 2B,rel})=(10^6\,M_\odot,0.1,10^{-4}\,{\rm yr}^{-1},10\,{\rm Gyr})$. Each grid point represents 300 Monte Carlo samples. The left and right panels show results for cases of $P_{\rm QPE}=1$ and 10 days, respectively. The region not sampled in our simulation is covered in light gray, and the region of TDE ($r_p<r_{p,\min}$) is colored in dark gray. The boundaries $t_{\rm EMRI}=3\,t_{\rm TDE}$ and $t_{\rm GW}=3\,t_{\rm TDE}$ are plotted in red solid and dashed lines. The typical eccentricity for bound stars produced from the Hills breakup, $e_{\rm Hills}$, is highlighted as the black dashed line.
    \label{fig:QPE-fraction-grid-fiducial}}
\end{figure*}

\subsubsection{QPE fraction}\label{sec:f-QPE-max-no-disk}

Our Monte Carlo model can in principle be combined with a realistic population of Hills breakup stars to forward predict the QPE rate.
However, the initial conditions of the Hills breakup stars depend on the binary fraction, binary SMA distributions, and relaxation time near the sphere of influence --- parameters that remain poorly constrained even in our Galactic Center \citep{Kocsis2011,Gautam2024}.
As an alternative, we explore the dependence of QPE orbit formation on model parameters by adopting the maximum fraction of QPE orbits on the Hills breakup line, $f_{\rm QPE,\max}$, as a proxy.

For a given MBH mass $\MBH$, TDE disk parameters $(h,\Gamma_{\rm TDE})$, and two-body relaxation time $t_{\rm 2B,rel}$, we consider 100 initial pericenter radii $r_{p,0}$ geometrically spaced over $[1.1,8]\times r_{p,\min}$. The initial eccentricity is set to $e_0=e_{\rm Hills}$, representative of orbits produced by Hills breakup. For each initial orbit $(a_0,r_{p,0})$, we generate 1000 Monte Carlo realizations. We then compute the fraction of stars that reach $P<P_{\rm QPE}$ for different period thresholds $P_{\rm QPE}$:
\begin{equation}
    f_{\rm QPE}(\MBH,h,\Gamma_{\rm TDE},t_{\rm 2B,rel},P_{\rm QPE},a_0,r_{p,0}).
\end{equation}
The maximum QPE fraction, $f_{\rm QPE,\max}$, is then defined as the largest value of $f_{\rm QPE}$ among these initial orbits:
\begin{multline}\label{eq:f-QPE-max}
f_{\rm QPE,\max}\lrb{\MBH,h,\Gamma_{\rm TDE},t_{\rm 2B,rel},P_{\rm QPE}}
=\\
\max_{e_0=e_{\rm Hills}}f_{\rm QPE}\lrb{\MBH,h,\Gamma_{\rm TDE},t_{\rm 2B,rel},P_{\rm QPE},a_0,r_{p,0}}.
\end{multline}

Figure~\ref{fig:f-QPE-max-no-disk} shows $f_{\rm QPE,\max}$ as a function of $\MBH$ for different $t_{\rm 2B,rel}$ in the absence of disk collisions ($\Gamma_{\rm TDE}=0\,{\rm yr}^{-1}$).
The maximum QPE fraction is negligible across all BH masses for $t_{\rm 2B,rel}=3\times10^9\,{\rm yr}$, with $f_{\rm QPE,\max}\lesssim0.01$, and remains small at $f_{\rm QPE,\max}\lesssim0.1$ for $t_{\rm 2B,rel}=10^{10}\,{\rm yr}$.
A significant fraction is only achieved when $t_{\rm 2B,rel}$ is increased to an unrealistically large value of $3\times10^{10}\,{\rm yr}$.
The need for such a long relaxation time to produce a significant QPE fraction through GW-driven orbital decay alone was also noted by \citet{Lu2023}.

Our Monte Carlo simulations and analytical argument demonstrate that for a two-body relaxation time less than a Hubble time ($t_{\rm 2B,rel}\lesssim10\,{\rm Gyr}$), GW emission is unable to overcome relaxation and the majority of stars produced from Hills breakup undergo tidal disruption.

\subsection{Interplay between GW, disk collisions, and relaxation}

We now use our Monte Carlo model to investigate how GW emission and star--disk collisions act together to overcome relaxation and produce QPE orbits. We first present the results of our fiducial model and then use $f_{\rm QPE,\max}$ (Eq.~\ref{eq:f-QPE-max}) to quantify the efficiency of QPE orbit production and examine its dependence on the model parameters.

\subsubsection{Fiducial model}

We choose $\MBH=10^6\,M_\odot$, $h=0.1$, $\Gamma_{\rm TDE}=10^{-4}\,{\rm yr}^{-1}$ and $t_{\rm 2B,rel}=10\,{\rm Gyr}$ as our fiducial model.
Figure~\ref{fig:QPE-fraction-grid-fiducial} shows $f_{\rm QPE}$ on a grid of initial orbits ($a_0,r_{p,0}$). With the inclusion of gas drag from disk collisions, a significantly higher fraction of stars can reach QPE orbits.
Compared to the case with only GW emission (Figure~\ref{fig:QPE-fraction-grid-no-disk}), the value of $f_{\rm QPE,\max}$ rises from $\sim5\%$ to $\sim30\%$.
This huge jump in QPE orbit formation probability can be understood by quantifying the efficiency of gas drag and GW emission in circularizing the orbit.
More specifically, we define an EMRI timescale that includes both circularization effects:
\begin{equation}\label{eq:t_EMRI}
    t_{\rm EMRI}=\frac{a}{\abs{(\Delta a)_{\rm disk}+(\Delta a)_{\rm GW}}}\cdot P.
\end{equation}
The boundary $t_{\rm EMRI}=3\,t_{\rm TDE}$, shown as the red solid line in Figure~\ref{fig:QPE-fraction-grid-fiducial}, reasonably captures our simulation results and confirms that this elevation of $f_{\rm QPE,\max}$ is a direct result of the increased circularization efficiency.

\subsubsection{Parameter dependence}

\begin{figure*}
    \centering
    \includegraphics[width=\textwidth]{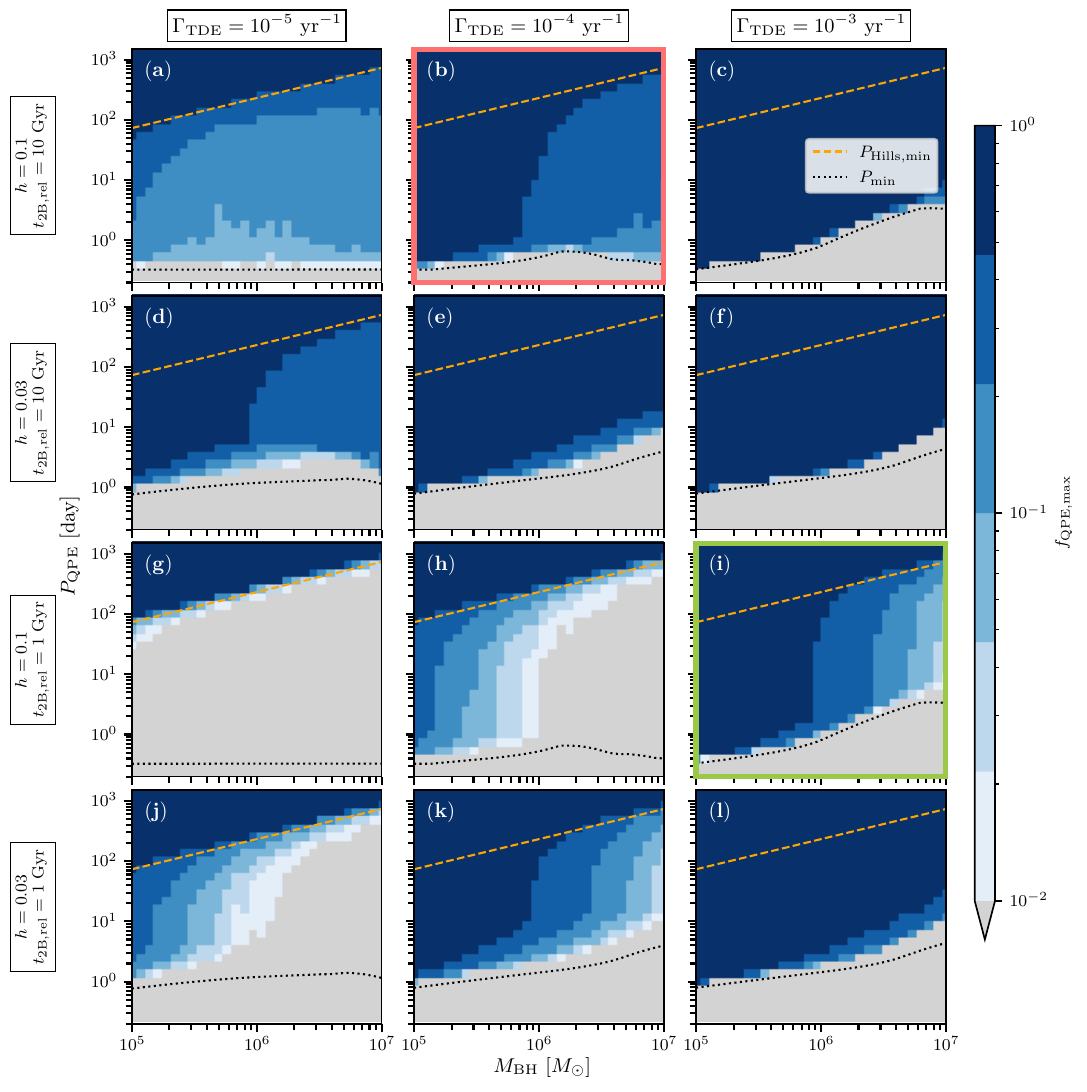}\\
    \caption{Maximum fraction of stars that form QPE orbits on the Hills breakup line $f_{\rm QPE,\max}$. Each row shows a combination of $h$ and $t_{\rm 2B,rel}$, and each column shows a choice of $\Gamma_{\rm TDE}$. Within each panel, we show $f_{\rm QPE,\max}$ as a function of $\MBH$ and $P_{\rm QPE}$. The semi-analytically computed $P_{\min}$ from the combination of $h$ and $\Gamma_{\rm TDE}$ is plotted as a black dotted line (see Figure~\ref{fig:P-min-GWdisk}), and the period minimum for Hills-captured stars is shown as a orange dashed line ($P_{\rm Hills,\min}$; Eq.~\ref{eq:P-Hills}). The red box highlights the fiducial model (panel b) and the green box highlights the preferred alternative model (panel i).
    \label{fig:f-QPE-max-grid}}
\end{figure*}

We now survey the dependence of QPE orbit formation efficiency on the model parameters with $f_{\rm QPE,\max}$ (Eq.~\ref{eq:f-QPE-max}). Figure~\ref{fig:f-QPE-max-grid} shows $f_{\rm QPE,\max}$ as a function of $(\MBH,P_{\rm QPE})$ for different combinations of $h,\,t_{\rm 2B,rel}$ (rows), and $\Gamma_{\rm TDE}$ (columns). In each panel, the orange line shows the minimum period expected from the Hills breakup:
\begin{multline}\label{eq:P-Hills}
    P_{\rm Hills,\min}=2\pi\sqrt{\frac{1}{G\MBH}\lrb{\frac{r_{p,\min}}{1-e_{\rm Hills}}}^3}\\=230\,{\rm days}\lrb{\frac{\MBH}{10^6\,M_\odot}}^{1/2}.
\end{multline}
The black curve shows the minimum period $P_{\min}$ calculated from the combination of $h$ and $\Gamma_{\rm TDE}$ (same as Figure~\ref{fig:P-min-GWdisk}).

We first focus on panel (b) that is highlighted in red. This panel has the same TDE disk parameters $(h=0.1,\,\Gamma_{\rm TDE}=10^{-4}\,{\rm yr}^{-1})$ and relaxation time $t_{\rm 2B,rel}=10\,{\rm Gyr}$ as our fiducial model. Across all BH masses, a large fraction of stars produced by Hills breakup can become EMRIs and reach sub-day period orbits. At higher $\MBH$, due to the higher initial eccentricity of the orbits and faster angular momentum relaxation, the value of $f_{\rm QPE,\max}$ drops slightly but remains above $0.1$. This is in stark contrast with the case without disk effects (Figure~\ref{fig:f-QPE-max-no-disk}), where $f_{\rm QPE,\max}<0.1$ for most BH masses and can only reach 0.1 at $\MBH=10^7\,M_\odot$.

We next reduce $\Gamma_{\rm TDE}$ to $10^{-5}\,{\rm yr}^{-1}$ while fixing $h$ and $t_{\rm 2B,rel}$ (panel a). At this lower TDE rate, $f_{\rm QPE,\max}$ drops for all BH masses but still remains higher than the case without disk effects. The decrease is the most significant for the intermediate BH mass of $\MBH\sim10^6\,M_\odot$, where neither gas drag from disk collision nor the GW orbital decay is sufficient to compete with relaxation. On the other hand, increasing $\Gamma_{\rm TDE}$ to $10^{-3}\,{\rm yr}^{-1}$ (panel c) leads to much higher $f_{\rm QPE,\max}$, especially for $\MBH\lesssim10^{6.5}\,M_\odot$. However, $f_{\rm QPE,\max}(P_{\rm QPE})$ also exhibits a sharp cutoff at a characteristic period that shifts to longer $P_{\rm QPE}$ with increasing $\MBH$. For $\MBH\gtrsim10^{6.5}\,M_\odot$, this cutoff prevents stars from reaching sub-day periods. This behavior reflects the period barrier imposed by the mass-loss efficiency (see \S\ref{sec:disk-only} and Eq.~\ref{eq:P-min-scaling}). The high TDE rate thus acts as a double-edged sword at $\MBH\gtrsim10^6\,M_\odot$ --- it enables efficient circularization that can compete with relaxation, but inhibits the formation of sub-day period orbits due to the accompanied mass loss. Sub-day QPE formation is most favorable at low BH masses $\MBH\lesssim10^6\,M_\odot$, and, at higher masses $\MBH\gtrsim10^6\,M_\odot$, for intermediate TDE rates of $10^{-5}\,{\rm yr}^{-1}\lesssim\Gamma_{\rm TDE}\lesssim10^{-4}\,{\rm yr}^{-1}$.

We next consider a thin disk with $h=0.03$ while keeping $t_{\rm 2B,rel}$ at 10\,Gyr (the second row in Figure~\ref{fig:f-QPE-max-grid}). Compared to $h=0.1$, a thinner disk is denser and can provide stronger gas drag to compete with relaxation. This leads to $f_{\rm QPE,\max}$ that is close to unity for $P_{\rm QPE}>10\,{\rm days}$. Nonetheless, a thinner disk also comes with higher mass loss efficiency $f_{\rm ml}$ (see Figure~\ref{fig:f-ml}), and prohibits stars from reaching sub-day periods across all considered $\MBH$. A thinner disk thus has effects on $f_{\rm QPE,\max}$ that are qualitatively similar to an increased $\Gamma_{\rm TDE}$.

For a reduced relaxation time of $t_{\rm 2B,rel}=1\,\rm Gyr$ (two bottom rows of Figure~\ref{fig:f-QPE-max-grid}), the value $f_{\rm QPE,\max}$ drops regardless of the choice of other parameters. This shows that relaxation is a strictly limiting factor and sets the timescale upper bound for any circularization mechanisms. The influence of a smaller $t_{\rm 2B,rel}$ is the most significant for $h=0.1$ and remains noticeable for $h=0.03$. For $h=0.1$, QPE orbit formation is completely stopped for $\Gamma_{\rm TDE}=10^{-5}\,{\rm yr}^{-1}$ across all BH masses, and the stars undergo tidal disruptions at the same periods as immediately after the Hills captures. At $\Gamma_{\rm TDE}=10^{-4}\,{\rm yr}^{-1}$, sub-day periods can only be reached at $\MBH\lesssim 10^{5.5}\,M_\odot$. The QPE formation at high BH masses $\MBH\gtrsim10^6\,M_\odot$ only becomes possible at the high TDE rate of $\Gamma_{\rm TDE}=10^{-3}\,{\rm yr}^{-1}$. For $h=0.03$, the more efficient circularization from the gas drag of the denser disks can better compete with relaxation, and the decrease in $f_{\rm QPE,\max}$ is less extreme.

Regardless of the disk thickness, a short relaxation time $t_{\rm 2B, rel}\lesssim 1\rm\, Gyr$ means that GW emission alone is insufficient to compete with relaxation, and the QPE formation relies on gas drag from disk collisions. However, since the final period that a star can reach only depends on the balance between circularization and mass loss efficiencies, a stronger gas drag only reinforces the period barrier (Figure~\ref{fig:P-min-GWdisk}), and orbits with sub-day periods at $\MBH\gtrsim10^6\,M_\odot$ become hard to achieve.

\subsubsection{Observations and favored models}
\begin{figure*}
    \centering
    \includegraphics[width=\textwidth]{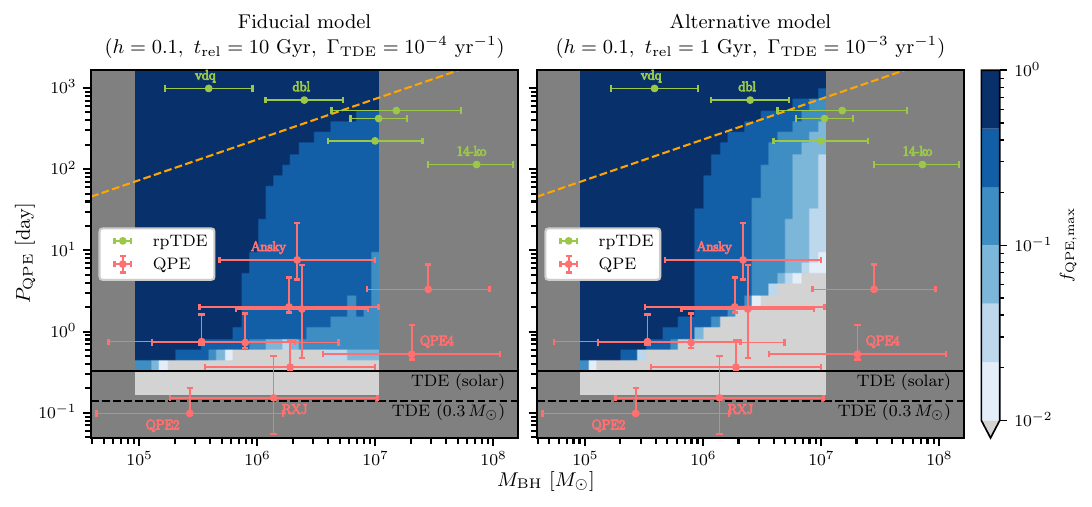}\\
    \caption{Maximum fraction of stars that form QPE orbits on the Hills breakup line $f_{\rm QPE,\max}$. The two panels show results for our fiducial model and favored alternative model, which are highlighted in Figure~\ref{fig:f-QPE-max-grid} with red and green frames (panel b and i). The light gray shade in the panel shows the region with $f_{\rm QPE,\max}<10^{-2}$, and the dark gray shade marks the region outside of our simulation. The orange line shows $P_{\rm Hills,\min}$ (Eq.~\ref{eq:P-Hills}), and the two black lines show the TDE period limit for a solar-mass star and a star with mass $M_*=0.3\,M_\odot$. In each panel, we also include the recurrence time of the known rpTDEs and QPEs. The  observational data are listed in Table~\ref{tab:recurrence-time}. For the QPE sources, we include an additional factor of two error in their period to accommodate the uncertain one or two-flare-per-orbit nature. There is also huge uncertainty on the $\MBH$ because many sources rely on the galaxy scaling relations that are poorly calibrated for low-mass galaxies \citep[see][]{Arcodia2025}.
    \label{fig:f-QPE-max-preferred}}
\end{figure*}

In Figure~\ref{fig:f-QPE-max-preferred} we show the comparison between observations and our preferred models. We choose two sets of model parameters --- one for our fiducial model ($h=0.1,t_{\rm 2B,rel}=10\,{\rm Gyr},\Gamma_{\rm TDE}=10^{-4}\,{\rm yr}^{-1}$; left panel) and one for the alternative model ($h=0.1,t_{\rm 2B,rel}=1\,{\rm Gyr},\Gamma_{\rm TDE}=10^{-3}\,{\rm yr}^{-1}$; right panel).
In both panels, we overlay the periods of the known repeating partial TDE (rpTDE) and QPE sources. The underlying data are listed in Table~\ref{tab:recurrence-time}. For QPE sources, we include an additional factor of two error bar in Figure~\ref{fig:f-QPE-max-preferred} to stress their unknown one- or two-flare-per-orbit nature. Some of the sources have BH mass estimates outside our simulation range $5<\log(\MBH/M_\odot)<7$. We do not attempt to simulate the system with $\MBH>10^7\,M_\odot$ because our model does not take into account the effects of general relativity on both the tidal disruption boundary and the TDE disk evolution at such high BH masses.

\begin{deluxetable*}{lccc}
\tabletypesize{\scriptsize}
\tablewidth{0pt}
\tablecaption{
Recurrence times and BH masses of QPEs and rpTDEs in the literature
\label{tab:recurrence-time}
}
\tablehead{
\colhead{Source} &
\colhead{Recurrence time or Period (day)} &
\colhead{$\log_{10}(M_{\rm BH}/M_\odot)$} &
\colhead{References}
}
\startdata
\sidehead{\textbf{QPEs}}
eRO-QPE1
& $0.73 \pm 0.11$
& $5.90 \pm 0.79$
& (a), (b)\\
eRO-QPE2
& $0.098 \pm 0.003$
& $5.43 \pm 0.79$
& (a), (b)\\
eRO-QPE3
& $0.75 \pm 0.06$
& $5.53 \pm 0.79$
& (b), (c)\\
eRO-QPE4
& $0.53 \pm 0.07$
& $7.31 \pm 0.75$
& (c)\\
eRO-QPE5
& $3.32 \pm 0.02$
& $7.45 \pm 0.52$
& (b)\\
GSN 069
& $0.37 \pm 0.02$
& $6.28 \pm 0.72$
& (b), (d)\\
RX J1301.9+2747
& $0.15 \pm 0.10$
& $6.14 \pm 0.88$
& (b), (e)\\
AT2019qiz
& $2.02 \pm 0.30$
& $6.27 \pm 0.76$
& (f)\\
AT2022upj
& $1.9 \pm 1.4$
& $6.38 \pm 0.56$
& (b), (g)\\
ZTF19acnskyy
& $7.6 \pm 3.3$
& $6.34 \pm 0.66$
& (b), (h), (i)\\
\sidehead{\textbf{rpTDEs}}
ASASSN-14ko
& $114$
& $7.86^{+0.31}_{-0.41}$
& (j)\\
eRASSt J045650.3$-$203750
& $223$
& $7.0 \pm 0.4$
& (k)\\
AT 2023adr
& $421$
& $7.03 \pm 0.24$
& (l)\\
AT 2023uqm
& $527$
& $7.18 \pm 0.55$
& (m)\\
AT 2022dbl
& $710$
& $6.40 \pm 0.33$
& (n)\\
AT 2020vdq
& $986$
& $5.59 \pm 0.37$
& (o), (p)\\
\enddata
\tablecomments{
QPE recurrence times are reported in the rest frame.
Owing to heterogeneous reporting in the source literature,
quoted uncertainties denote the reported spread or range
and do not necessarily represent $1\sigma$ errors.
References:
(a) \citet{Arcodia2021};
(b) \citet{Arcodia2025};
(c) \citet{Arcodia2024a};
(d) \citet{Miniutti2019};
(e) \citet{Giustini2024};
(f) \citet{Nicholl2024};
(g) \citet{Chakraborty2025};
(h) \citet{Hernandez-Garcia2025a};
(i) \citet{Hernandez-Garcia2025b};
(j) \citet{Payne2021};
(k) \citet{Liu2023};
(l) \citet{Angus2026};
(m) \citet{Wang2026};
(n) \citet{Lin2024};
(o) \citet{Yao2023};
(p) \citet{Somalwar2025}.
}
\end{deluxetable*}

Most rpTDE sources, except for AT 2022dbl and AT 2020vdq, have periods below $P_{\rm Hills,\min}$ and likely have experienced a certain degree of circularization to evolve to their current orbits. 
Even though we do not explicitly model the orbits with BH mass $\MBH>10^7\,M_\odot$, the extrapolation of our fiducial model suggests that the gas drag from disk collisions should be sufficient for their orbital circularization.

The situation for QPE sources is more complicated. Although our fiducial model can reliably produce QPEs with recurrence time $t_{\rm rec}>0.5\,{\rm day}$, it struggles to produce the sources with $t_{\rm rec}<0.5\,{\rm day}$ due to the period barrier. For eRO-QPE2 and RX J1301.9+2747, their $t_{\rm rec}$ are so much lower than our $P_{\min}$ that even demanding two QPE flares per orbit ($P_{\rm orb}=2\,t_{\rm rec}$) cannot explain the discrepancy relative to our fiducial models. One additional challenge in producing orbits for these two sources is that they are below the lower period limit allowed by the tidal disruption of solar-like stars.
% In Figure~\ref{fig:f-QPE-max-preferred}, we plot the period limit of tidal disruption using $a=2\,r_{t}$.
For this reason, we show in Figure~\ref{fig:f-QPE-max-preferred} the period limits of tidal disruption $P(a = 2\,r_t)$ for a solar-type star and an M-dwarf, with $M_*=0.3\,M_\odot$ and $R_*=0.38\,R_\odot$ following the scaling $R_*\propto M_*^{0.8}$. Explaining the orbits of eRO-QPE2 and RX J1301.9+2747 would require an M-dwarf or stripped star that is more resilient to tidal disruption and likely two QPE flares per orbit.

Another potential issue with our fiducial model is that it requires both a long relaxation time $t_{\rm 2B,rel}=10\,\rm Gyr$ and a moderately high TDE rate $\Gamma_{\rm TDE}=10^{-4}\,{\rm yr}^{-1}$. However, an elevated TDE rate is likely accompanied by a boosted rate of Hills breakup that will increase the number density of stars with $P\lesssim\,$a few days. This may reduce the relaxation time and make $t_{\rm 2B,rel}=10\,\rm Gyr$ hard to achieve. An alternative model, shown in the right panel of Figure~\ref{fig:f-QPE-max-preferred}, has a higher TDE rate $\Gamma_{\rm TDE}=10^{-3}\,{\rm yr}^{-1}$ and a shorter relaxation time $t_{\rm 2B,rel}=1\,\rm Gyr$. This model can produce QPEs with $t_{\rm rec}\gtrsim1\,\rm day$, but does not produce QPEs with shorter periods and likely cannot explain rpTDEs like ASASSN-14ko if it indeed has $\MBH \sim 10^8M_\odot$. This model is also potentially in tension with eRO-QPE4, which has a relatively short $t_{\rm rec}$ with a likely high $\MBH$. However, the large uncertainties in these $\MBH$ estimates (based on host galaxy scaling relations) make it difficult to assess the significance of this tension \citep{Arcodia2025,Wevers2026}.

\section{Summary and Discussion}
\label{sec:discussion}
In this paper we have followed the long-term, coupled orbital and stellar evolution of Hills-captured Sun-like stars around $10^5$--$10^7\,M_\odot$ MBHs, asking under what conditions they can reach the short-period, low-eccentricity orbits required by orbiter models of QPEs. Our framework starts with modeling the coupled evolution of the star and disk, including gas drag from collisions with recurrent, viscously spreading TDE disks, per-collision mass loss calibrated to the hydrodynamic simulations of \citet{Yao2025},  and the star's \texttt{MESA}-based mass--radius response (\S\ref{sec:gas-drag-mass-loss}\,\&\,\ref{sec:disk-only}). We then incorporate additional physical processes in successive sections: GW-driven orbital decay in \S\ref{sec:disk-GW-model} and stochastic two-body relaxation of the orbital angular momentum, implemented through a Monte Carlo model, in \S\ref{sec:full-model}.

Our main conclusions can be summarized as follows.

{\bf First, drag from TDE disk collisions efficiently circularizes orbits, but stellar ablation imposes a minimum period.} Because the drag and the mass loss both scale linearly with the disk surface density, this period floor is insensitive to how much gas the star encounters and is instead set by the intrinsic physics of the collisions; it scales approximately as $P_{\rm min} \propto M_{\rm BH}$ at high black hole masses, where it also becomes independent of the disk aspect ratio and excludes sub-day orbits for $M_{\rm BH} \gtrsim 10^{6}\,M_\odot$ from disk drag alone (\S\ref{sec:disk-only}).

{\bf Second, the relative importance of GW decay compared to disk drag increases with $\MBH$ for typical Hills-capture orbits (\S\ref{sec:circularization-timescales}; Figure~\ref{fig:t_drag-t_GW-ratio}):} disk drag dominates circularization around low-mass MBHs while GW emission takes over for $M_{\rm BH} \gtrsim 10^{6\text{--}7}\,M_\odot$. Once GW orbital decay is included, sub-day periods become accessible across the full MBH mass range for $h = 0.1$ and $\Gamma_{\rm TDE} \lesssim 10^{-4}\,\rm yr^{-1}$.

{\bf Third, the conventional EMRI-formation criterion, $t_{\rm GW} = t_{L,\rm rel}$, overestimates the circularization fraction of Hills-captured stars:} their proximity to the loss cone means tidal disruption typically precedes circularization even within the nominally GW-dominated region \citep{Kaur2025}.
A boundary based instead on the tidal disruption timescale, $t_{\rm GW}=3t_{\rm TDE}$ (Eq.~\ref{eq:GW-rel-bd-new}), matches our Monte Carlo results and remains accurate with disk drag when $t_{\rm GW}$ is replaced by the combined circularization time $t_{\rm EMRI}$ (Eq. \ref{eq:t_EMRI}). This confirms that disk drag increases the maximum fraction of Hills-captured stars reaching QPE orbits, $f_{\rm QPE,max}$, by accelerating circularization.
For two-body relaxation times below a Hubble time, GW decay alone leaves few survivors, with $f_{\rm QPE,max} \lesssim 0.01$ for $t_{\rm 2B,rel} = 3$ Gyr and $\lesssim 0.1$ for $t_{\rm 2B,rel} = 10$ Gyr (\S\ref{sec:full-model}). Episodic TDE-disk drag is therefore the decisive ingredient: for a 10-day period threshold, $h=0.1$, and $t_{\rm 2B,rel}=10\,\mathrm{Gyr}$, it raises $f_{\rm QPE,max}$ to $\sim0.1$ at $\Gamma_{\rm TDE}=10^{-5}\,{\rm yr}^{-1}$, to $\sim0.3$ in our fiducial model ($\Gamma_{\rm TDE}=10^{-4}\,{\rm yr}^{-1}$), and toward unity at still higher TDE rates (Figure~\ref{fig:f-QPE-max-grid}). The preference of QPEs for high TDE rates may explain why they are preferentially found in post-starburst galaxies, where TDE rates are enhanced \citep{Wevers2022,Wevers2024,Hammerstein2021,Yao2023,Shepherd2026}.

{\bf Fourth, formation of sub-day QPEs around $M_{\rm BH} \gtrsim 10^{6}\,M_\odot$ favors a TDE rate of $10^{-5}\,\rm yr^{-1}\lesssim\Gamma_{\rm TDE} \lesssim10^{-4}\,\rm yr^{-1}$ and a relaxation time of $t_{\rm 2B,rel}\gtrsim10\,{\rm Gyr}$ (Figure~\ref{fig:f-QPE-max-grid}).} This combination allows circularization to outpace relaxation while retaining a significant contribution from GW emission, which lowers the period minimum to less than a day (Figure~\ref{fig:P-min-GWdisk}). In contrast, gas drag around $M_{\rm BH} \lesssim 10^{6}\,M_\odot$ is more efficient and can bring the stars to sub-day periods even without GW emission. Sub-day QPEs in low-mass MBHs thus have no upper limit on the TDE rate and can tolerate faster relaxation. This preference for low $\MBH$ is qualitatively consistent with the observed QPE sample, in which the shortest-period sources are hosted by the least massive black holes \citep{Arcodia2021,Giustini2020,Giustini2024,Wevers2022,Wevers2024}.

{\bf Fifth, comparison with observations reveals two parameter regimes, each with distinct strengths and limitations (\S\ref{sec:full-model}; Figure~\ref{fig:f-QPE-max-preferred}).}
Except for AT 2022dbl and AT 2020vdq, all rpTDEs in our compilation lie below $P_{\rm Hills,\min}$ and require circularization, which
extrapolation of our fiducial model above $10^7\,M_\odot$ suggests disk drag can provide. The fiducial model also reproduces QPEs with
$t_{\rm rec}\gtrsim0.5\,{\rm day}$, but its period floor excludes eRO-QPE2 and RX J1301.9+2747, even if two flares are emitted per orbit. Both also lie below the tidal-disruption limit of a solar-type star, and require a more resilient M-dwarf or stripped star and two flares per orbit.
Despite these successes, the fiducial model requires $t_{\rm 2B,rel}=10\,{\rm Gyr}$ and $\Gamma_{\rm TDE}=10^{-4}\,{\rm yr^{-1}}$, a potentially inconsistent combination:
the elevated TDE rate implies more frequent Hills captures, increasing the density of short-period stars and shortening the relaxation time.
The alternative model ($t_{\rm 2B,rel}=1\,{\rm Gyr}$, $\Gamma_{\rm TDE}=10^{-3}\,{\rm yr^{-1}}$) avoids this tension and reproduces QPEs with $t_{\rm rec}\gtrsim1\,{\rm day}$, but fails to produce shorter periods or high-$\MBH$ rpTDEs.
eRO-QPE4 poses the strongest potential tension: its short recurrence time and high inferred $\MBH$ cannot be reproduced with $t_{\rm 2B,rel}=1\,{\rm Gyr}$, although the large uncertainty in its BH mass estimate makes this tension difficult to assess \citep{Arcodia2025,Wevers2026}.

{\bf Sixth, sub-day orbits require geometrically thick TDE disks, $h \sim 0.1$:} for $h = 0.03$, nearly all stars reach 10-day orbits but sub-day periods are suppressed (Figure \ref{fig:f-QPE-max-grid}). 
At sufficiently short orbital periods, the disk height satisfies $H/R_*\lesssim 70$, and the mass loss per collision is governed by the peak ram pressure, $\Delta M_*\propto p_{\rm ram}\propto \Sigma/h$ (Eq.~\ref{eq:mass-loss-general}). Thus, at fixed surface density, a thicker disk produces the same orbital drag ($\Delta E_{\rm drag}\propto \Sigma$; Eq.~\ref{eq:Delta-E-drag}) but less mass loss, allowing the star to survive to shorter orbital periods. Maintaining an aspect ratio $h \sim 0.1$ as the accretion rate falls well below Eddington is difficult with thermal pressure support, but is qualitatively consistent with magnetically elevated disks supported by toroidal fields \citep{Begelman2007, Jiang2014, Sadowski2016, Begelman2017, Curd2019, Kaur2023}. If sub-day QPEs form through this channel, they therefore point to magnetically supported TDE disks. 

Taken together, these results support the basic viability of the orbiter model: the combination of Hills capture, recurrent TDE disks, and GW decay can deliver surviving stars to QPE-like orbits with order-unity efficiency in plausible regions of parameter space, while simultaneously predicting a period floor, a preference among sub-day systems for lower-mass MBHs, and a tight link between QPE activity and the nuclear TDE history. The sources that resist this picture (eRO-QPE2, RXJ1301, and eRO-QPE4) are informative: the first two point toward low-mass or stripped orbiters, and the last toward a long relaxation time in at least some high-mass nuclei.

These conclusions rest on a simplified description of the TDE disk evolution. We model each disk as a one-zone, constant-aspect-ratio $\alpha$-disk that spreads self-similarly, with an initial mass of $0.13$--$0.5\,M_\odot$ set by Eddington-limited fallback and a maximum extent of $10\,r_{d,0}$. The masses, spreading rates, and thicknesses of real TDE disks remain debated: late-time UV photometry has been interpreted as evidence for compact, slowly evolving disks \citep{vanVelzen2019}, recent late-time SED modeling favors disks extending to roughly ten times the circularization radius \citep{Guolo2025}, and theoretical expectations for the spreading rate span a considerable range \citep{Cannizzo1990, ShenMatzner2014, Mummery2020}. The vertical structure, meanwhile, may be controlled by magnetic elevation rather than thermal pressure \citep{Begelman2007, Kaur2023}, which a constant $h$ captures only crudely. Encouragingly, the period floor is insensitive to the disk surface density; the main sensitivities are to the aspect ratio --- thinner disks strengthen the drag but also raise the period floor, so that for $h = 0.03$ nearly all stars reach 10-day orbits while sub-day systems are suppressed except around the lowest-mass black holes --- and to the maximum disk extent $r_d(t_f)$, which sets the range of pericenters that experience drag at all.

The mass-loss prescription carries comparable uncertainty. The per-collision stripping is taken from the fitting formulae of \citet{Yao2025}, which were calibrated against 2D \texttt{Athena++} simulations of a polytropic star over a limited range of ram pressures and disk scale heights, and are extrapolated here across a much wider dynamic range. The transition between the single- and multiple-shock regimes --- governed by the ratio of the shocked envelope's thermal time to the orbital period, and treated here simply by taking the maximum of the two prescriptions --- is uncertain at the factor-of-a-few level. The uncertainty in the mass loss efficiency propagates into the limiting period as $P_{\rm final}\propto f_{\rm ml}^{3/2}$. The simulations also do not account for realistic orientations between the stellar orbit and the disk plane and neglect the effects of tidal heating \citep{Yao2026}.

Our treatment of relaxation is similarly idealized. We adopt a single, radius-independent two-body relaxation time and follow only the diffusion of angular momentum, yet $t_{\rm 2B,rel}$ near the sphere of influence is poorly constrained even in the Galactic Center, varies with the structure of the nuclear star cluster, and can be effectively shortened by massive perturbers and resonant relaxation \citep{RauchTremaine1996}. Since $f_{\rm QPE,max}$ changes by more than an order of magnitude between $t_{\rm 2B,rel} = 1$ and $10$ Gyr, this is the dominant systematic uncertainty in our survival fractions, and one that likely varies substantially from nucleus to nucleus.

Our model also does not include stellar collisions, which could either boost or hinder QPE formation. If stars are destroyed during  collisions, the population of stars in short-period orbit and the QPE rate would be suppressed. On the other hand, if the stellar cores survive, the substantial energy dissipation during collisions could lead to a rapid episode of orbital circularization. Debris produced by stellar collisions could also circularize into an accretion disk and exert gas drag on the surviving stellar cores, analogous to the TDE disks considered in our model.
Properly modeling the collision rate and outcomes is beyond the scope of the current work.

These uncertainties compound when converting survival fractions into predicted QPE rates. Such a forward model requires the Hills capture rate, and hence the binary fraction and separation distribution near the sphere of influence \citep{Hills1988, Yu2003, Zhang2013} --- quantities that are poorly constrained observationally --- as well as processes omitted from our model: in dense nuclei, direct star--star collisions can strip or destroy stars on highly eccentric orbits on timescales competitive with those of the circularization processes considered here \citep{Rose2024}. We therefore regard a detailed comparison with the observed QPE abundance \citep{Arcodia2024b} as premature, although the order-unity QPE fractions achieved for $\Gamma_{\rm TDE} \sim 10^{-4}\,\rm yr^{-1}$ are qualitatively consistent with scenarios in which QPEs accompany a non-negligible fraction of TDEs around low-mass MBHs \citep{Linial2023}.

Despite these caveats, the recurrent spreading-disk picture makes testable predictions. The emergence of QPEs four years after the TDE AT 2019qiz \citep{Nicholl2024} places that system in the earliest phase of our disk evolution. Because the local surface density at the orbiter's radius declines steeply ($\Sigma \propto t^{-4/3}$) once the disk has engulfed the orbit, the model predicts that the flares should grow secularly fainter over a timescale of a couple of years as the disk drains, with QPE activity eventually shutting off until a subsequent TDE replenishes the gas. However, this prediction does not take into account the fate of stripped stellar material, which could secularly replenish the accretion disk, hence helping to maintain the QPE flare luminosity \citep{Lu2023}. Continued X-ray monitoring of AT 2019qiz and other post-TDE QPE systems can therefore shed light on the validity of this framework.

Finally, although our integrations terminate when $M_* < 0.4\,M_\odot$ or $r_{\rm p} < 2\,r_{t,\odot}$, the subsequent evolution carries observable consequences. Once the star is stripped down to its nearly isentropic core, its mass--radius relation inverts and the radius grows with further mass loss (\S\ref{sec:mass-loss}); the enlarged cross section then increases both the per-collision stripping and the flare energetics, suggesting a runaway in which the eruptions grow progressively brighter as the star is destroyed --- the likely fate of the longer-period, more common systems. For the rarer short-period systems near the period floor, the combination of shrinking mass, growing radius, and hence growing tidal radius may instead bring the star into Roche-lobe contact, transitioning the system from collision-powered eruptions to tidal mass transfer onto the MBH \citep{linialSari2023, Lu2023, Yao2026}. Distinguishing these endpoints, brightening flares versus the onset of mass transfer, may become possible with the long observational baselines now accumulating for the oldest known QPE sources.

\section*{Acknowledgments}

We thank Joheen Chakraborty, Itai Linial, and Brenna Mockler for insightful conversations. HH's and WL's research was supported by an LSST Scialog Early Science grant from the Research Corporation for Science Advancement and by a Sloan Research Fellowship (Award Number FG-2026-79505) from the Alfred P. Sloan Foundation.  This work was supported by the Gordon and Betty Moore Foundation through grant GBMF 5076.01.

\bibliographystyle{mnras}

% You should give the same name for your .bbl as your main .tex
% since it is a requirement for posting on ArXiv.
\bibliography{main}

\end{document}